\documentclass[twocolumn,showpacs,preprintnumbers,amsmath,amssymb]{revtex4}
\usepackage{amsfonts,diagbox,latexsym,amssymb,graphics,epsfig,subfigure,color,makeidx}
\usepackage{xcolor}
\usepackage{array}
\usepackage{booktabs}
\usepackage{makecell} 
\usepackage{amsmath}
\usepackage[utf8]{inputenc}
\usepackage{graphicx,hyperref}
\usepackage{subcaption}
\usepackage{caption}
\usepackage{bm}
\usepackage{color}
\usepackage{multirow}

\newcommand{\dd}{{\rm d}}

\begin{document}
	\title{Shadows of quintessence black holes: spherical accretion, photon trajectories, and geodesic observers}
	\author{Ji-Wen Li$^{1}$, 
		Zi-Liang Wang$^{1,2}$ \footnote{ziliang.wang@just.edu.cn}, and Tao-Tao Sui$^{3}$
        \footnote{suitaotao@aust.edu.cn}
	}
	\affiliation{$^{1}$ Department of Physics, School of Science, Jiangsu University of Science and Technology,
Zhenjiang, 212003, People's Republic of China\\
		$^{2}$ Theoretical Physics Research Center, School of Science, Jiangsu University of Science and Technology, Zhenjiang, 212003, People's Republic of China\\
        $^{3}$ Center for Fundamental Physics,
School of Mechanics and Photoelectric Physics,
Anhui University of Science and Technology,
Huainan, 232001, People's Republic of China}
	
\begin{abstract}
The presence of a quintessence-like field can influence the black hole shadow through three primary mechanisms: the dynamics of accretion flows, the trajectories of photons, and the motion of observers. 
Unlike standard shadow analyses that assume a static observer at spatial infinity, the non-asymptotically flat nature of quintessence-corrected spacetimes motivates the consideration of freely falling (geodesic) observers. 
Using a perturbative approach, we derive analytical expressions for the event-horizon location, photon-sphere radius, innermost stable circular orbit, and critical impact parameter. We compute the observed intensity profiles for both static and infalling spherical accretion flows. We find that, although the photon-sphere radius and the critical impact parameter are invariant properties of the spacetime, the apparent angular size of the shadow depends sensitively on the observer’s motion and location. Freely infalling observers systematically measure smaller angular radii than static observers at the same radius, whereas freely outgoing observers measure larger ones, in agreement with relativistic aberration. 
In contrast to the Schwarzschild case, the impact parameter alone is insufficient to characterize the observed angular structure in non-asymptotically flat spacetimes. Applying our results to the Event Horizon Telescope observation of M87$^\ast$, we show that more negative equations of state lead to stronger constraints on the quintessence parameter, largely independent of the observer prescription.  Our analysis highlights the importance of carefully specifying the observer in shadow studies of non-asymptotically flat black-hole spacetimes.

\end{abstract}

	
	\maketitle
	
	\section{INTRODUCTION}

Since the first images of the supermassive black holes M87$^\ast$~\cite{EventHorizonTelescope:2019dse,EventHorizonTelescope:2019ggy} and Sgr A*~\cite{EventHorizonTelescope:2022wkp} were released by the Event Horizon Telescope (EHT), the morphology of black hole shadows has become an essential tool for testing gravity theories~\cite{Vagnozzi:2022moj} and probing the astrophysical environments surrounding black holes. Motivated by these achievements, numerous studies have explored how quantum effects, scalar hair, plasma effects, and cosmological backgrounds influence the shadow geometry~\cite{Gralla:2019xty,Ozel:2021ayr,Bisnovatyi-Kogan:2019wdd,Wang:2023rjl,Chang:2019vni,Meng:2025ivb,Chen:2024nua,Sui:2023yay,Gao:2023mjb,Hou:2022eev,Zhang:2024lsf}.

In the presence of a quintessential field, Kiselev~\cite{Kiselev:2002dx} derived a class of spherically symmetric black hole solutions by solving Einstein’s field equations with an energy-momentum tensor corresponding to a surrounding quintessence-like field. Here, by ``quintessence-like" we emphasize that Kiselev's quintessence is not a perfect fluid, unlike the quintessence commonly considered in cosmology~\cite{Zlatev:1998tr,Tsujikawa:2013fta,Sahni:2002kh,Visser:2019brz}.
 The resulting metric generalizes the Schwarzschild spacetime, with the line element (in natural units)
\begin{align}
    \dd s^2 = -f(r)\, \dd t^2 + \frac{1}{f(r)}\, \dd r^2 + r^2 \left( \dd \theta^2 + \sin^2 \theta\, \dd \phi^2 \right),
\end{align}
where
\begin{equation}
    f(r) = 1 - \frac{2M}{r} - \sum_n \left( \frac{r_n}{r}\right)^{3\omega_n + 1}.
\end{equation}
Here, $ M $ denotes the black hole mass, $ \omega_n $ are the equation-of-state parameters of different components, and $ r_n $ are the corresponding dimensional normalization constants. For $-1<\omega_n< -1/3 $, this family of ``quintessence black holes’’ provides an effective phenomenological framework linking local gravitational physics with large-scale cosmological effects.  Quintessence black holes have been studied from various perspectives, including their geodesic structure, black hole shadow, and thermodynamic properties~\cite{Fernando:2012ue,Zhang:2006ij,Thomas:2012zzc,Ghaderi:2016dpi,Chen:2005qh,Zeng:2020vsj,Belhaj:2020rdb,Sun:2022wya,
Mustafa:2022xod,Atamurotov:2022nim,Singh:2023zmy,Hamil:2023zeb,Ahmed:2025boj}. Although these spacetimes are static and cannot, in general, be rewritten in the McVittie form~\cite{McVittie:1933zz}--which describes a central mass embedded in an expanding universe--they still capture the essential features of a black hole surrounded by a dark-energy-like background.

Most previous investigations of shadows cast by quintessence black holes have focused on
static observers located at finite distances.
These studies provide valuable insights into how the presence of quintessence
modifies the optical appearance of black holes and offer a useful reference for comparison. Nevertheless, it should be noted that static observers may not always provide a fully reliable
physical description, especially in quintessence-corrected spacetimes that are not
asymptotically flat.
A well-known example is the Schwarzschild-de~Sitter spacetime, which can be understood
as a Schwarzschild black hole embedded in a de~Sitter universe and corresponds to an
extreme case of quintessence with $\omega = -1$.
As shown in Ref.~\cite{Perlick:2018iye}, the angular radius of the  shadow for the Schwarzschild--de~Sitter black hole
shrinks to a finite, nonvanishing value when a comoving observer approaches infinity.

Therefore, in non-asymptotically flat spacetimes, the notion of a static observer at large distances
may lose a clear physical meaning.
In contrast, geodesic observers--freely falling test particles following timelike geodesics--provide a more consistent and physically well-defined description of local measurements.
Their local rest frames determine how photons are received and, consequently, the apparent shape of the black hole shadow.

In this work, we investigate how the presence of a quintessence-like field affects the black hole shadow through three primary mechanisms: the dynamics of accretion flows, the trajectories of photons, and the motion of observers. 
Specifically, we study the impact of the quintessential field on spherical accretion and examine the resulting optical appearance of the accretion flow. 
We analyze how quintessence-induced modifications of the spacetime geometry influence both null and timelike geodesics. The former governs photon propagation, while the latter determines the motion of observers. 

Particular attention is paid to the shadow size as perceived by different classes of observers, including static observers and geodesic observers. Finally, we discuss the asymptotic appearance of the black hole shadow at large observer distances and comment on possible constraints on the quintessence parameter from EHT observations.

This paper is organized as follows. 
In Section~\ref{sec:II}, we review the Schwarzschild black hole with a quintessence correction and introduce a perturbative method. 
Within this framework, we derive approximate analytical expressions for the event horizon radius, the photon sphere radius, the innermost stable circular orbit, and the critical impact parameter. 
In Section~\ref{sec:sec3}, we investigate the optical appearance of the quintessence-corrected Schwarzschild black hole as perceived by a static observer, taking into account both static and radially infalling spherical accretion flows. 
In Section~\ref{sec:sec4}, we extend our analysis to a geodesic (freely falling) observer and focus on the apparent size of the photon ring, both within the domain of outer communication and in the region outside the outer horizon. 
Finally, Section~\ref{Conclusion} is devoted to conclusions and discussion.

Throughout this paper, we work in natural units, setting the speed of light to unity and taking $G=1$, where $G$ denotes Newton’s gravitational constant.

\section{Quintessence Black Hole: Horizons, photon sphere, ISCO and radial geodesic}\label{sec:II}

In the original construction by Kiselev, the surrounding matter distribution can in principle be described by a superposition of several components with different equations of state. 
In this paper, however, we focus on a simplified and physically transparent scenario in which the spacetime metric is dominated by a single quintessence-like field characterized by a constant equation-of-state parameter. 
This assumption is adopted for two main reasons. 
First, it allows us to isolate and systematically analyze the influence of a given $\omega$ on the spacetime geometry and, consequently, on the black hole shadow. 
Second, at a given radial scale, it is reasonable to expect that one effective quintessence component provides the leading contribution, while subdominant components can be neglected to a good approximation. 

Under these assumptions, we consider a static and spherically symmetric spacetime describing a black hole surrounded by a minimally modeled quintessence field,
\begin{equation}
    \dd s^2 = -f(r)\dd t^2 + \frac{1}{f(r)}\dd r^2 + r^2\left(\dd \theta^2 + \sin^2\theta \dd \phi^2\right),
    \label{eq:metric}
\end{equation}
where the metric function is given by
\begin{equation}
    f(r) = 1 - \frac{2M}{r} - \frac{c}{r^{3\omega+1}} .
\end{equation}
Here, $c$ is a non-negative normalization constant, and $\omega$ is the equation-of-state parameter of the fluid. For a quintessence field, the physically relevant range is $-1 < \omega < -\tfrac{1}{3}$. For simplicity, we have suppressed dimensional considerations in the above expressions. 
Since the combination $c / r^{3\omega+1}$ is dimensionless, the radial coordinate in this term should, strictly speaking, be divided by a characteristic length scale $r_d$. 
In other words, the expression implicitly takes the form
\begin{align}
    \frac{c\, r_d^{3\omega+1}}{r^{3\omega+1}}, 
\end{align}
which renders $c$ dimensionless. 
Throughout the remainder of this paper, we set $r_d = M$. 
This choice does not introduce any artificial restriction, as $c$ itself is an arbitrary dimensionless parameter.

In order to determine the event horizons, one must solve $f(r)=0$, that is,
\begin{equation}
    1 - \frac{2M}{r} - \frac{c}{r^{3\omega+1}} = 0.
\end{equation}
Multiplying both sides by $r^{3\omega+1}$ gives
\begin{equation}\label{eq:eom_rh}
    r^{3\omega+1} - 2M r^{3\omega} - c = 0.
\end{equation}
In general, this equation does not admit a closed analytic solution for arbitrary values of $\omega$.
However, for several special cases--corresponding to integer or simple fractional powers of $r$--the horizons can be obtained analytically.

When $\omega = -\tfrac{2}{3}$, the equation reduces to a quadratic form,
\begin{equation}
    c\,r^{2} - r + 2M = 0,
\end{equation}
which yields two analytic roots,
\begin{equation}
    r_{\pm} = \frac{1 \pm \sqrt{1 - 8 M c}}{2c},
\end{equation}
representing the inner (black hole) and outer horizons, respectively. 

In the extreme case  of $\omega = -1$, the metric reduces to the Schwarzschild–de~Sitter case,
\begin{equation}
    1 - \frac{2M}{r} - c r^{2} = 0,
\end{equation}
and the three real roots of this cubic equation can be written in terms of Cardano’s formula, yielding one negative and two positive horizons corresponding to the black hole and cosmological horizon.

For the other values in the range $-1 < \omega < -\tfrac{1}{3}$, there are, in general, two horizons.
The inner horizon typically corresponds to the black-hole horizon, while the outer horizon at a larger radius plays a role similar to a cosmological horizon. Following Ref.~\cite{Perlick:2018iye}, we refer to the region between these two horizons as the domain of outer communication, since it corresponds to the portion of the spacetime that is causally connected to distant observers and allows for two-way communication without crossing either horizon.

Within this parameter range, Eq.~\eqref{eq:eom_rh} involves non-integer powers of $ r $ and therefore does not admit a closed-form analytic solution. 
Consequently, the horizon radii can in general only be determined numerically. 
Nevertheless, for physically relevant regimes, we show that approximate analytical solutions can be obtained within a perturbative framework.

In realistic astrophysical situations, the quintessence correction to the Schwarzschild metric has a significant influence only at very large radii. Therefore, the term ${c}/{r^{3\omega+1}}$ can be regarded as a small perturbation near the black hole scale, i.e., for $r \sim M$. Assuming that the black-hole horizon can be written as~\cite{Wang:2025fmz}
\begin{align}\label{eq:r_h_ass}
    r_\text{h} = R_\text{S}\left[1+  f_1 \left(\frac{c}{R_\text{S}^{3\omega+1}}\right)\right],
\end{align}
where $R_{\rm S} = 2M$ is the Schwarzschild radius and the dimensionless coefficient $f_1$ will be determined shortly.  
Substituting Eq.~\eqref{eq:r_h_ass} into Eq.~\eqref{eq:eom_rh} and keeping only the leading order in $c/R_{\rm S}^{3\omega+1}$, one obtains $f_1 = 1$, and therefore
\begin{align}\label{eq:r_h_sol}
    r_\text{h} = R_\text{S} + \frac{c}{R_\text{S}^{3\omega}}.
\end{align}

We emphasize that the perturbative expansion presented above is intended to compute the correction to the black-hole event horizon at the local scale $r \sim 2M$, rather than the outer horizon.  
Nevertheless, we will see that it can also be used to derive an approximate expression for circular orbits around the black hole.

Due to the spherical symmetry of spacetime, particle's motion can be confined to the equatorial plane, i.e., $\theta=\pi/2$. In this plane, two conserved quantities along the geodesic (with affine parameter $\lambda$) can be defined ~\cite{Wald}:
\begin{equation}
    E=f(r)\frac{dt}{d\lambda}\label{eq:energy},
\end{equation}
\begin{equation}
    L=r^2\frac{d\phi }{d\lambda}\label{eq:momentum}.
\end{equation}
Substituting Eqs.~\eqref{eq:energy} and ~\eqref{eq:momentum} into the four-velocity normalization condition,
\begin{equation}
    g_{\mu \nu}\frac{dx^\mu}{d\lambda} \frac{dx^\nu}{d\lambda}=-N,
\end{equation}
we obtain
\begin{equation}
    \left(\frac{dr}{d\lambda}\right)^2+\left(\frac{L^2}{r^2}+N\right)f(r)=E^2\label{eq:velocity},
\end{equation}
with the constant $N=0$ for a massless particle (null geodesic) and $N=1$ for a massive particle (timelike geodesic). Eq.~\eqref{eq:velocity} can be written as 
\begin{equation}
   \frac{1}{2}\left(\frac{dr}{d\lambda}\right)^2+V_{\text{eff}}=\frac{E^2}{2}\label{eq:rsolve}, 
\end{equation}
where the effective potential is defined as
\begin{equation}
    V_{\text{eff}}=\frac{f(r)}{2}\left(\frac{L^2}{r^2}+N\right).\label{eq:potential}
\end{equation}

\begin{table*}[htbp]
\centering
\renewcommand{\arraystretch}{1.5} 
\begin{tabular}{|c|ccc|ccc|}
\hline
\multirow{2}{*}{} & \multicolumn{3}{c|}{$\omega=-1/2$} & \multicolumn{3}{c|}{$\omega=-3/4$} \\
\cline{2-7}
 & \makebox[2.4cm]{$c=10^{-3}$} & \makebox[2.4cm]{$c=10^{-4}$} & \makebox[2.4cm]{$c=10^{-5}$} & \makebox[2.4cm]{$c=10^{-3}$} & \makebox[2.4cm]{$c=10^{-4}$} & \makebox[2.4cm]{$c=10^{-5}$} \\
\hline
$r_\text{h}$ numerical & 2.00283 & 2.00028 & 2.00000 & 2.00478 & 2.00048 & 2.00005 \\
$r_\text{h}$ analytical & 2.00283 & 2.00028 & 2.00003 & 2.00476 & 2.00048 & 2.00005 \\
\hline
$r_\text{{ps}}$ numerical & 3.00390 & 3.00039 & 3.00000 & 3.00446 & 3.00044 & 3.00005 \\
$r_\text{{ps}}$ analytical & 3.00390 & 3.00039 & 3.00004 & 3.00444 & 3.00044 & 3.00004 \\
\hline
$r_\text{{ISCO}}$ numerical & 6.02229 & 6.00221 & 6.00022 & 6.29459 & 6.02371 & 6.00233 \\
$r_\text{{ISCO}}$ analytical & 6.02205 & 6.00220 & 6.00022 & 6.23242 & 6.02324 & 6.00232 \\
\hline
$b_\text{c}$ numerical & 5.20969 & 5.19750 & 5.19629 & 5.22718 & 5.19923 & 5.19646 \\
$b_\text{c}$ analytical & 5.20969 & 5.19750 & 5.19629 & 5.22718 & 5.19923 & 5.19646 \\
\hline
\end{tabular}
\caption{Comparison between numerical and analytical results for the event horizon radius ($r_{\rm h}$), the photon sphere radius ($r_{\rm ps}$), the innermost stable circular orbit ($r_{\rm ISCO}$), and the critical impact parameter ($b_{\rm c}$) of quintessence-corrected Schwarzschild black holes for different values of the parameters $\omega$ and $c$.}
\label{tab:comparison}
\end{table*}

For null geodesics $(N=0)$, the effective potential is given by
\begin{equation}
    V_{\text{null}}=\frac{L^2f(r)}{2r^2}\label{eq:null}.
\end{equation}
The radius of the photon sphere, denoted as $r_\text{ps}$, is determined by the condition $\partial V_{\text{null}}/\partial r=0$. For a Schwarzschild black hole surrounded by a quintessence field, this leads to the following equation:
\begin{equation}
   -2r^{3\omega+1}+6M\,r^{3\omega}+3c(\omega+1)=0\label{eq:r_sp}.
\end{equation}
Similarly, we assume that the radius of the corrected photon sphere is given by
\begin{equation}
    r_\text{ps}=R_\text{ps}\left[1+f_{2}\left(\frac{c}{R_\text{ps}^{3\omega+1}}\right)\right]\label{eq:r_s},
\end{equation}
where $R_{\rm ps}=3M$ is the radius of the photon sphere for the Schwarzschild black hole. Substituting Eq.~\eqref{eq:r_s} into Eq.~\eqref{eq:r_sp} and retaining only the leading order in $c/R_{\rm ps}^{3\omega+1}$, we can obtain
\begin{equation}
    f_2=\frac{3(\omega +1)}{2},
\end{equation}
therefore, 
\begin{equation}\label{eq:ps}
    r_\text{ps}=R_\text{ps}+\frac{3(\omega +1)}{2}\frac{c}{R_\text{ps}^{3\omega}}.
\end{equation}

For $\omega=-1$, Eq.~\eqref{eq:ps} reduces to $r_\text{ps}=R_\text{ps}$, which agrees with the result provided in Ref.~\cite{Perlick:2018iye}.

We now turn to timelike geodesics. 
In this case, the effective potential derived from Eq.~\eqref{eq:potential} takes the form
\begin{equation}
    V_{\text{t}}=\frac{f(r)}{2}\left(\frac{L^2}{r^2}+1\right)\label{eq:timelike}.
\end{equation}
The radius of innermost stable circular orbit(ISCO) should satisfy $\partial V_{\text{t}}/\partial r=\partial^2 V_{\text{t}}/\partial^2 r=0$, from which one can obtain 
\begin{align}\label{eq:eom_r_{ISCO}}
0=&c r^{3 \omega} \left[6 M (\omega (3 \omega-4)-2)-9 r \omega^2+r\right] \notag \\ 
&-3 c^2 (\omega+1) (3 \omega+1) -2 M (6 M-r) r^{6 \omega}.
\end{align}
One can assume the following form for the corrected ISCO radius:
\begin{equation}\label{eq:assum_ISCO}
    r_{\rm ISCO} =R_{\rm ISCO}\left[ 1+f_3\left( \frac{c}{R_{\rm ISCO}^{3\omega+1}}\right)\right],
\end{equation}
where $R_{\text{ISCO}}=6M$ is the radius of the ISCO for
the Schwarzschild black hole. Substituting Eq.~\eqref{eq:assum_ISCO} into Eq.~\eqref{eq:eom_r_{ISCO}} and retaining only the leading order correction, we have 
\begin{align}
    f_3=3(6\omega^2+4\omega+1),
\end{align}
thus,
\begin{equation}
    r_{\rm ISCO}=R_{\rm ISCO}+3(6\omega^2+4\omega+1)\left(\frac{c}{R_{\rm ISCO}^{3\omega}}\right).
\end{equation}

So far, we have derived perturbative analytical solutions for circular orbits of both massive and massless particles. 
We now turn to the critical impact parameter of the quintessence-corrected Schwarzschild black hole, which determines whether photons are ultimately captured by the black hole or escape to infinity. 
The impact parameter is defined as $ b \equiv L/E $, and the critical impact parameter for photons is obtained from the condition $ V_{\rm null}(r_{\rm ps}) = E^2/2 $. 
This yields
\begin{equation}
    \label{eq:b_cr}
    b_{\rm c} = \frac{r_{\rm ps}}{\sqrt{f(r_{\rm ps})}} \, .
\end{equation}

Table~\ref{tab:comparison} presents a comparison of the event-horizon radius $r_{\rm h}$, the photon-sphere radius $r_{\rm ps}$, the innermost stable circular orbit $r_{\rm ISCO}$, and the critical impact parameter $b_{\rm c}$ as obtained from numerical calculations and from the perturbative analytical solutions of the Schwarzschild metric with a single quintessence correction.
The close agreement between the two sets of results demonstrates that the perturbative approach provides a valid and physically meaningful description in the vicinity of the black hole.



\section{Shadow Observed by a Static Observer}
\label{sec:sec3}
\begin{figure*}
    \centering
    \includegraphics[width=0.7\textwidth]{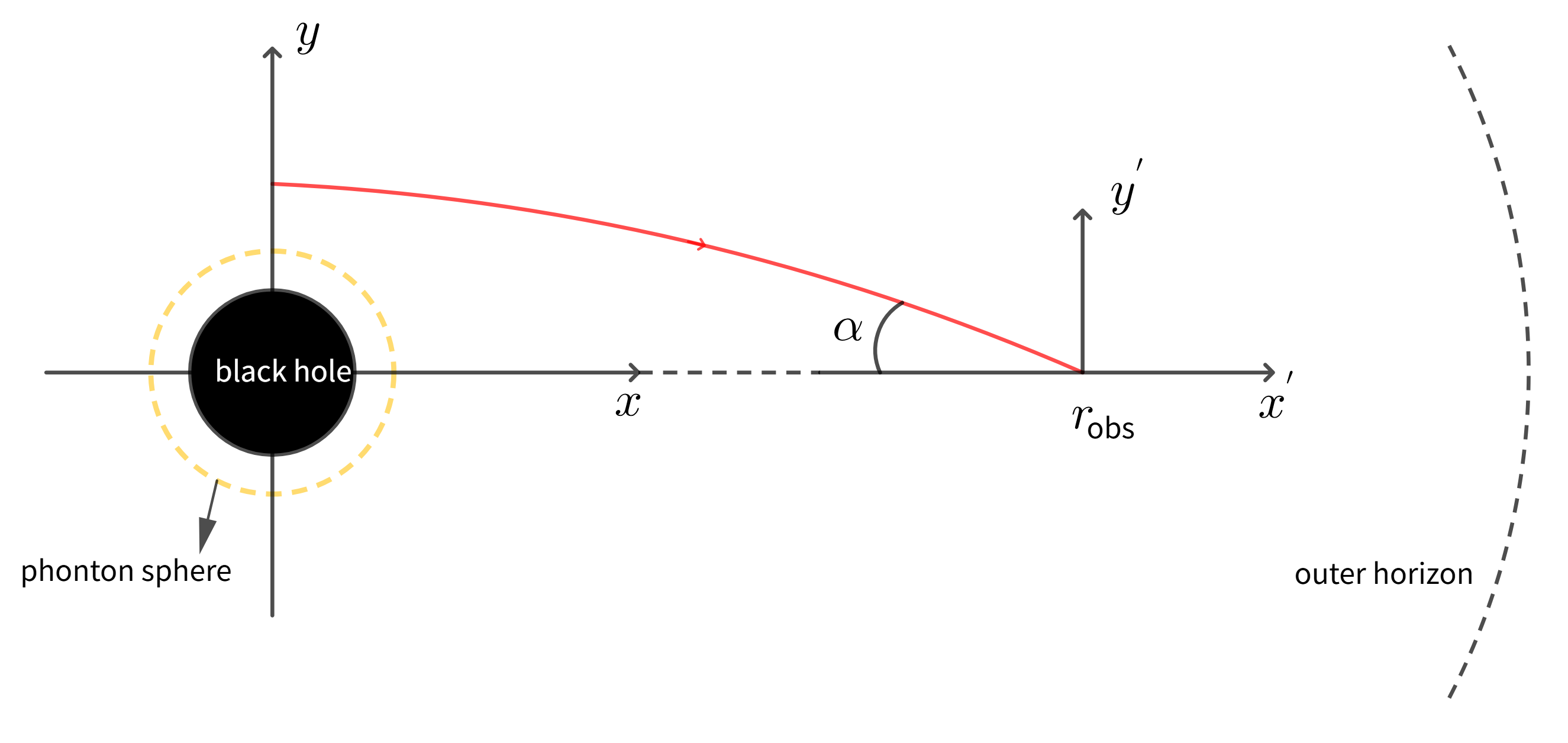} 
    \caption{Coordinate systems for a photon trajectory (red solid curve) in the equatorial plane ($\theta = \pi/2$). 
The Euclidean coordinates $(x, y)$ are defined as $x = r \cos\phi$ and $y = r \sin\phi$. 
The $(x', y')$ coordinates represent the local reference frame of an observer located at radius $r = r_{\rm obs}$. 
The center of the black hole is located at the origin of the Euclidean coordinate system $(x, y)$.}
    \label{fig:tetrad}
\end{figure*}

In order to calculate the black hole shadow in different reference frames, 
it is convenient to introduce the tetrad formalism. 
An observer's local proper reference frame $({t}', {x}', {y}', {z}')$ 
is constructed using a non-coordinate orthonormal basis via~\cite{Carroll2004}:
\begin{subequations}
\label{eq:tetrads}
\begin{align}
     \frac{\partial}{\partial {x'}^{a}} 
    = e_{a}{}^{\mu}\,\frac{\partial}{\partial x^{\mu}},
\end{align}
with the dual basis satisfying
\begin{align}
    \dd {x'}^{a} = e^{a}{}_{\mu}\, \dd x^{\mu},
\end{align}
\end{subequations}
where Latin indices $a,b,\dots$ denote frame components ($x'^{0}=t'$, $x'^{1}=x'$, $x'^{2}=-z'$, and $x'^{3}=y'$), 
while Greek indices $\mu,\nu,\dots$ correspond to coordinate components. 
The quantities $e_{a}{}^{\mu}$ and $e^{a}{}_{\mu}$ are inverse matrices, 
commonly referred to as tetrads (or vielbeins). 
They form a $4\times4$ matrix with positive determinant and satisfy the orthonormality condition
\begin{align}\label{eq:orthonormality_condition}
    g_{\mu\nu}\, e_{a}{}^{\mu} e_{b}{}^{\nu} = \eta_{ab},
\end{align}
as well as the completeness relation
\begin{align}
    g_{\mu\nu} = e^{a}{}_{\mu} e^{b}{}_{\nu}\, \eta_{ab},
\end{align}
where $\eta_{ab}$ denotes the Minkowski metric. 
In asymptotically flat spacetimes, observers are usually assumed to be static at spatial infinity, where the spacetime metric approaches the Minkowski form. However, for a black hole surrounded by a quintessence field, the geometry is no longer asymptotically flat, and the concept of an observer ``at infinity'' becomes ill-defined. 
Instead, we could introduce a static observer located at a finite radial coordinate $r_{\text{obs}}$. 
This observer remains at rest with respect to the coordinate system and measures the direction of incoming photons in their local orthonormal reference frame (see Fig.~\ref{fig:tetrad}).  In this case, the corresponding orthonormal tetrad is given by 
\begin{align}
    e^{a}{}_{\mu}= \mathrm{diag}\left( \sqrt{f(r)},\,1/ \sqrt{f(r)},\, r,\, r\sin\theta \right)\,.
\end{align}
For an incoming light ray originating in the vicinity of the black hole (see, e.g., the red curve in Fig.~\ref{fig:tetrad}), its apparent angle $\alpha$ in the local reference frame, as measured by a static observer located at $r_{\rm obs}$, is given (for $\theta=\pi/2$) by
\begin{align}
\tan \alpha 
&= \mp \left.\frac{\dd y'}{\dd x'} \right|_{r_{\rm obs}} \notag\\
&= \mp \left.\frac{e^{3}{}_{3}\,\dd \phi}{e^{1}{}_{1}\,\dd r}\right|_{r_{\rm obs}} \notag\\
&= \pm \left.\frac{\sqrt{f(r)}}{r}
\left(\frac{E^{2}}{L^{2}} - \frac{f(r)}{r^{2}}\right)^{-1/2}\right|_{r_{\rm obs}}\,,
\label{eq:tana}
\end{align}
where the upper sign corresponds to photons arriving from the negative $x'$-axis direction (as illustrated in Fig.~\ref{fig:tetrad}), whereas the lower sign corresponds to photons arriving from the positive $x'$-axis direction. The latter situation can occur, for example, when the observer is located inside the photon sphere.

From Eq.~\eqref{eq:tana}, one can obtain that 
\begin{align}
    \sin ^2 \alpha = \frac{b^2 f(r_{\rm obs})}{r_{\rm obs}^2}\label{eq:angle}\,.
\end{align}
For asymptotically flat spacetimes, one has $ f(r_{\rm obs}) \to 1 $ for a fixed distant observer. 
In this limit, Eq.~\eqref{eq:angle} reduces to $ \alpha \simeq b / r_{\rm obs} $, so that the impact parameter $ b $ is directly proportional to the observed angular size. 
As a consequence, $ b $ is often used as a convenient proxy for the apparent size when analyzing the black hole shadow~\cite{Gralla:2019xty,Wang:2025czc}. 

In non-asymptotically flat spacetimes, however, this identification is no longer generally valid. 
The truly observable quantity is the angle $ \alpha $, defined with respect to the observer’s local orthonormal frame, whereas the impact parameter $ b $ is a conserved quantity along null geodesics and does not, by itself, encode the local measurement process. 
This distinction is sometimes overlooked in the literature, where $ b $ is implicitly interpreted as the apparent angular size even in non-asymptotically flat backgrounds, which may lead to misleading conclusions.

This issue can be illustrated by considering the behavior at the horizons, where $ f(r)=0 $. 
Eq.~\eqref{eq:angle} then immediately implies $ \sin\alpha = 0 $ for static observers located at the horizons. 
Strictly speaking, no timelike observer can remain static on a horizon; nevertheless, the limiting behavior of static observers approaching the horizon can still be meaningfully considered. 
In particular, one finds
\begin{align}
\alpha = 0 \,, \qquad &\text{at the outer horizon}, \notag\\
\alpha = \pi \,, \qquad &\text{at the inner (black-hole) horizon}.
\end{align}
Physically, this behavior is understood as follows.  For a static observer hovering just around the outer horizon, the outward-pointing photon cone becomes tangent to the radial direction.  
As a consequence, the observer sees all allowed photon directions compressed into a vanishing solid angle around the outward radial direction, so the ``apparent size'' of the black hole shrinks to zero.  
In other words, a static observer at the outer horizon would find that the black hole has no finite angular size in the sky.

By contrast, a static observer located at the inner horizon finds the situation reversed.  Here the roles of the ingoing and outgoing photon cones interchange, and the only available photon directions point toward decreasing $r$.  The angular coordinate $\alpha = \pi$ indicates that the entire visible sky is filled by the black-hole region. Thus the black hole covers the full $4\pi$ steradians of the observer’s sky, and the ``apparent size'' becomes the entire celestial sphere.

  \begin{figure*}[htbp]
    \centering
    \includegraphics[width=0.48\linewidth]{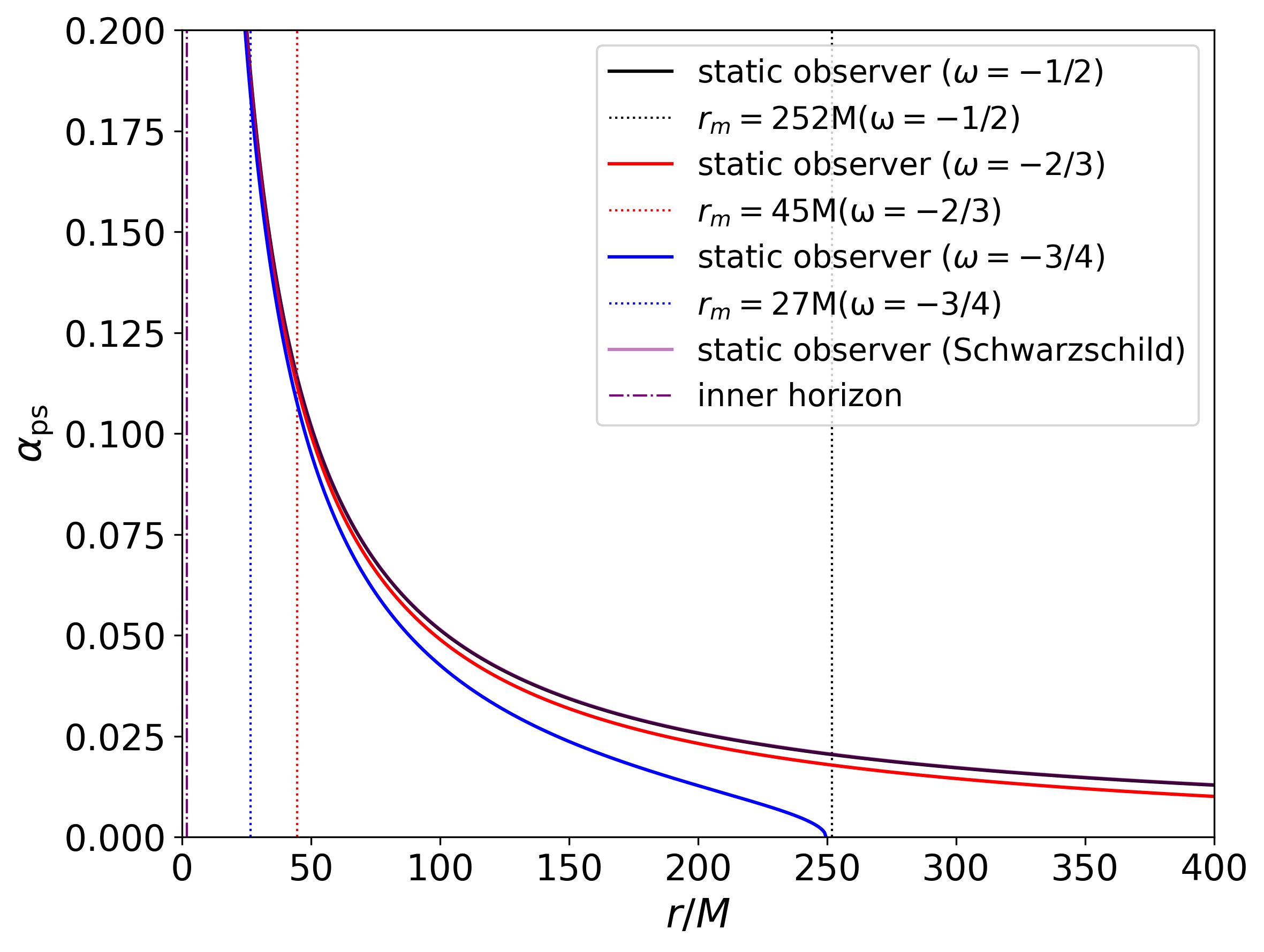}
    \hfill
    \includegraphics[width=0.48\linewidth]{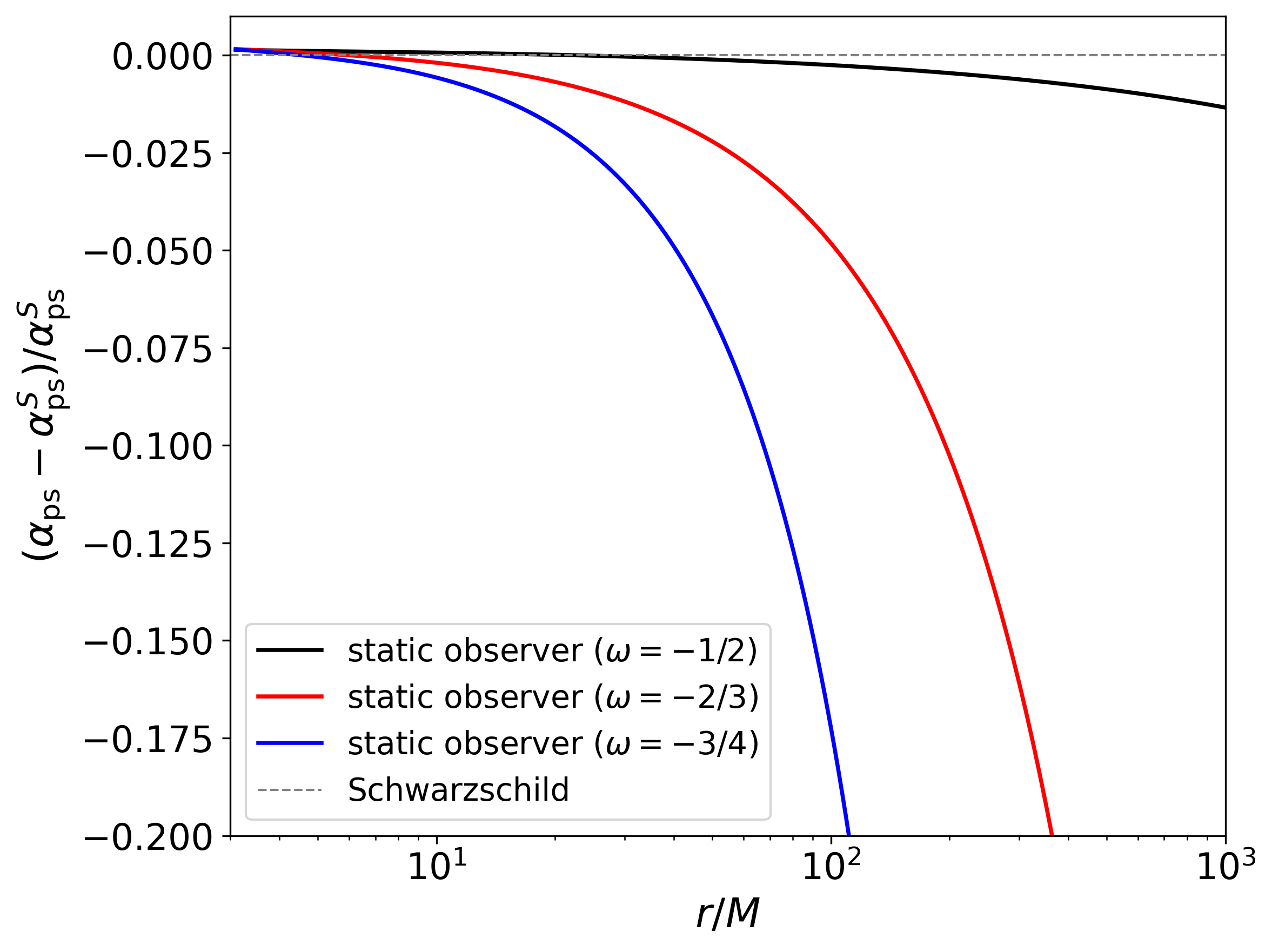}
    \caption{Left panel: Angular radius of the photon sphere observed by static observers at different radii for several black-hole spacetimes.  For a given value of $\omega$, only the observer located at $r_{\rm m}$ follows a geodesic; all other static observers have nonzero proper acceleration.  
Note that the curve for Schwarzschild case almost overlaps with the black curve.  Right panel: Comparison of the photon-sphere angular radius observed by static observers in quintessence black-hole spacetimes and in the Schwarzschild spacetime.  
For all quintessence black holes, the normalization constant is fixed to $c=0.001$.
}
\label{fig:static_comparing}
\end{figure*}

These two limiting behaviors further demonstrate that the apparent angular size of the black hole varies continuously from zero at the outer horizon to the full sky at the inner horizon.  For a static observer located at the photon sphere, substituting the critical impact parameter given in Eq.~\eqref{eq:b_cr} into Eq.~\eqref{eq:angle}, we find that the apparent angular radius of the photon sphere is $\pi/2$. A plot of the angular radius of the photon sphere, as observed by static observers at different radii, is shown in Fig.~\ref{fig:static_comparing} (the definition of $r_{\rm m}$ in the figure will be clarified later).

We now examine the influence of the quintessence field on accretion flows. To this end, we consider a simple spherically symmetric accretion model and assume at first that the radiating gas is at rest in the static frame of the spacetime.  This idealized setup allows us to isolate the effect of the metric modification on the emitted radiation, in close analogy with the treatment adopted in Ref.~\cite{Narayan:2019imo}.
 
Since the gas is at rest in the static frame, its specific energy is entirely  gravitational and is given by $E(r)=\sqrt{f(r)}$.  
Therefore, the total gravitational energy that can be released as the matter is moved quasi–statically from $r_1$ to a smaller radius $r$ is simply the 
difference
\begin{align}\label{eq:energydiff}
    \Delta E = \sqrt{f(r_{1})} - \sqrt{f(r)}\,.
\end{align}

Assuming that all the released energy is converted into radiation, a convenient way to express the energy difference is through the integral
\begin{equation}
\begin{aligned}
   \Delta E 
   &= \int_{r}^{r_{1}} 
      f(r')\, P_{\mathrm{t}}(r')\, 4\pi r'^{\,2}\,
      \frac{1}{\sqrt{f(r')}}\, \mathrm{d}r'  \\
   &= \sqrt{f(r_1)} - \sqrt{f(r)},
\end{aligned}
\end{equation}
where the factor $f(r')$ accounts for the redshift of local energy and time intervals, and $1/\sqrt{f(r')}$ converts proper radial distance to coordinate radius~\cite{Narayan:2019imo}. Here,
\begin{equation}
    P_\mathrm{t}(r)
    = \frac{f^{\prime}(r)}{8\pi r^{2} f(r)}
\end{equation}
is the total radiated power per unit proper volume.  

Assuming isotropic emission, the (bolometric) total emission coefficient per unit solid angle 
in this model can be written as
\begin{equation}
    j_\mathrm{t}(r) = \frac{P_\mathrm{t}(r)}{4\pi}
                     = \,\frac{f^{\prime}(r)}{32 \pi^2 r^2 f(r)}\,.
    \label{eq:emission}
\end{equation} 

In the absence of any quintessence corrections, Eq.~\eqref{eq:emission} 
reduces to the emission coefficient per unit solid angle for a Schwarzschild 
black hole, as derived in Ref.~\cite{Narayan:2019imo}:
\begin{equation}
    j_\mathrm{S}(r) = \frac{M}{16 \pi^2 r^4 \left(1 - \frac{2M}{r}\right)}.
\end{equation}
While this expression reproduces the standard Schwarzschild treatment, our formulation is in fact more general: it applies to any static and spherically symmetric metric, therefore remains valid even in non–asymptotically flat spacetimes such as those considered in this work. This generality allows us to systematically isolate the effects of the modified metric functions on the gravitational redshift, the energy release, and the observed radiation.

\begin{figure}
    \centering
    \includegraphics[width=0.9\linewidth]{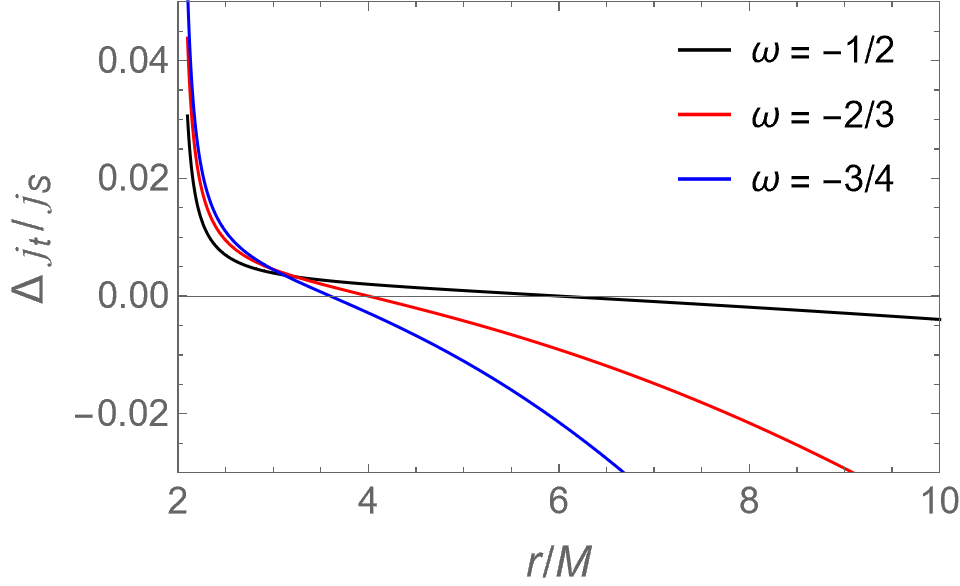}
    \caption{Difference in the total emission coefficient, $\Delta j_{\rm t} = j_{\rm t} - j_{\rm S}$, between the
quintessence–corrected Schwarzschild spacetime and the standard Schwarzschild case. For all choices of the equation-of-state parameter $\omega$, the
normalization constant is fixed to $c=0.001$.}
    \label{jt_difference}
\end{figure}

In Fig.~\ref{jt_difference}, we show the difference in the total emission 
coefficient between the quintessence–modified spacetime and the standard 
Schwarzschild case, defined as 
$\Delta j_{\rm t} = j_{\rm t} - j_{\rm S}$.
For a fixed normalization constant $c$, it is evident that the deviation becomes increasingly pronounced as the equation-of-state parameter $\omega$ takes more negative values. For a given $\omega$ in Fig.~\ref{jt_difference}, the emission coefficient in the quintessence-modified spacetime is smaller than in the Schwarzschild case at large radii but becomes larger at smaller radii. This behaviour arises from the competition between $f'(r)$ and $f(r)$ in the expression for the emission coefficient given in Eq.~\ref{eq:emission}, which leads to different scaling properties across radius in the two spacetimes.

In the theory of radiative transfer, a useful quantity is the invariant intensity, defined as~\cite{mihalas1984foundations}
\begin{equation}
    \mathcal{I}_\nu \equiv \frac{I_\nu}{\nu^3},
\end{equation}
where $I_\nu$ is the specific intensity and $\nu$ is the photon frequency. 
The invariant intensity $\mathcal{I}_\nu$ satisfies the relativistic radiative transfer equation~\cite{mihalas1984foundations}:
\begin{equation}
    \frac{d\mathcal{I}_\nu}{d\lambda}
    = \frac{j(\nu)}{\nu^2}
    - \nu\,\mathcal{I}_\nu\,\chi(\nu),
    \label{eq:transfer}
\end{equation}
where $j(\nu)$ is the specific emission coefficient and $\chi(\nu)$ is the specific absorption coefficient.
The specific emission coefficient is related to the total emission coefficient in Eq.~\eqref{eq:emission} through
\begin{align}
    j_{\rm t} = \int j(\nu)\, d\nu\, .
\end{align}
For simplicity, we neglect all absorption processes, which is appropriate for the optically thin regime relevant to EHT observations.

According to Eq.~\eqref{eq:transfer} the observed intensity can be obtained by integrating the emissivity along the light ray
\begin{equation}
    I_\text{obs}(\nu_\text{obs},r_\text{obs})=\nu^3_\text{obs}\int_\text{ray}\frac{j(\nu_\text{em})}{\nu^2_\text{em}} \mathrm{d}\lambda \label{eq:intensityv}.
\end{equation}
The photon frequency measured by an observer is given by 
\begin{equation}
    \nu=-g_{\alpha\beta}k^\beta u^\alpha\label{eq:frequency},
\end{equation}
where $u^\alpha$ is the 4-velocity of an observer and $k^\beta$ is the 4-momentum of the photon. Thus, the redshift factor is given by
\begin{equation}
    g=\frac{\nu_\text{obs}}{\nu_\text{em}}=\frac{k_\alpha u^\alpha _\text{obs}}{k_\beta u^\beta _\text{em}}\label{eq:redshift},
\end{equation}
where $u^\beta _\text{em}$ is the 4-velocity of the emitter.

Then the observed intensity
can be obtained by integrating all observed frequencies
\begin{equation}
    I_\text{obs}=-\int_\text{ray} \frac{j_\text{t}g^3k_{\alpha}u^\alpha_\text{obs}}{k^r}\, \mathrm{d}r\label{eq:all},
\end{equation}
where $k^{r}\equiv{\mathrm{d}r}/{\mathrm{d}\lambda}$. 

Several remarks are in order regarding our approach. 
First, although $j_{\mathrm t}$ is originally derived for a static, rest–radiating gas, 
we argue that it may also be used for an infalling emitter in the context of our simplified model. 
For an infalling gas, the relevant energy difference is no longer given by Eq.~\eqref{eq:energydiff}; 
instead, it may be roughly proportional to
 $\sqrt{f(r_{\mathrm m})} - \sqrt{f(r)} $.
Such a modification changes the observed intensity profile only by an overall multiplicative factor, 
without altering its radial dependence in a way that is important for our purposes. 
Therefore, the qualitative image morphology remains essentially unchanged. In the literature (e.g. Ref.~\cite{Narayan:2019imo}), this substitution is typically adopted directly. 
Here, we provide a brief justification for its validity and then consider two types of emitter motion in our analysis: 
a static (rest–radiating) model and an almost free-fall model.  In both cases, we employ the same total emission coefficient $j_{\mathrm t}$,  interpreted as the emissivity measured in the local comoving rest frame of the gas.

Second, Eq.~\eqref{eq:all} can be integrated numerically in a direct manner for a  given emitter–observer pair, without performing full geodesic ray tracing. 
This feature differs from the method employed in Ref.~\cite{Narayan:2019imo}. 

\begin{figure}
    \centering
    \includegraphics[width=0.95\linewidth]{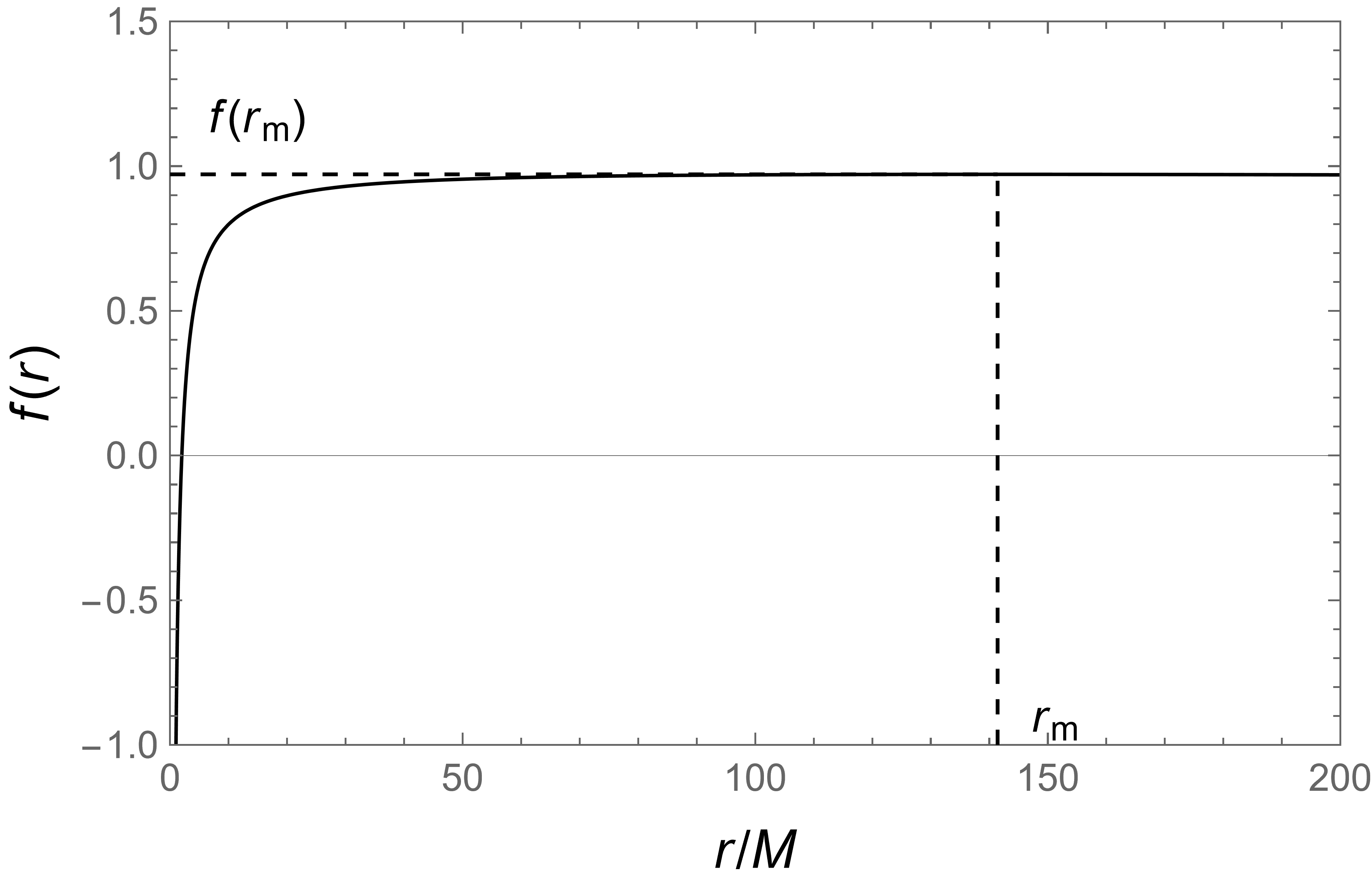}
\caption{Metric time component $f(r)$ of the quintessence-corrected Schwarzschild black hole for the equation-of-state parameter $\omega=-2/3$ and normalization constant $c=10^{-4}$. The function $f(r)$ attains its maximum at $r_{\rm m}$.}
    \label{fig:time component}
\end{figure}

Third, the choice of a static observer requires particular care.  In the Schwarzschild spacetime it is customary to place the observer at spatial infinity, where the metric function approaches unity, $\displaystyle \lim_{r\to\infty} f(r) = 1$.  
For a quintessence--corrected black hole, however, the spacetime is generally not asymptotically flat, and no canonical notion of a static observer at infinity exists. A physically well-motivated choice is to place the observer at the radius 
where $f(r)$ attains its maximum value,
   $ r_{\rm obs} = r_{\rm m}$, see Fig.~\ref{fig:time component}. 
At this location we have, $f'(r_{\rm m})=0$, 
and thus a static worldline has vanishing radial proper acceleration.  
Consequently, the static observer at $r_{\rm m}$ is momentarily following a 
timelike geodesic\footnote{This equilibrium is unstable, but the local inertial  interpretation remains valid for our purpose.}.   This provides a natural extension of the usual asymptotic observer to non--asymptotically flat, static spherically symmetric spacetimes. Nevertheless, in the following discussion we will also consider alternative choices of static observers located at $r_{\rm obs} \neq r_{\rm m}$ when needed for specific purposes. In such cases, the observer possesses a nonzero proper acceleration and therefore follows a nongeodesic trajectory.

\subsection{Static Spherical Emission}
    
\begin{figure}
    \centering
    \includegraphics[width=0.9\linewidth]{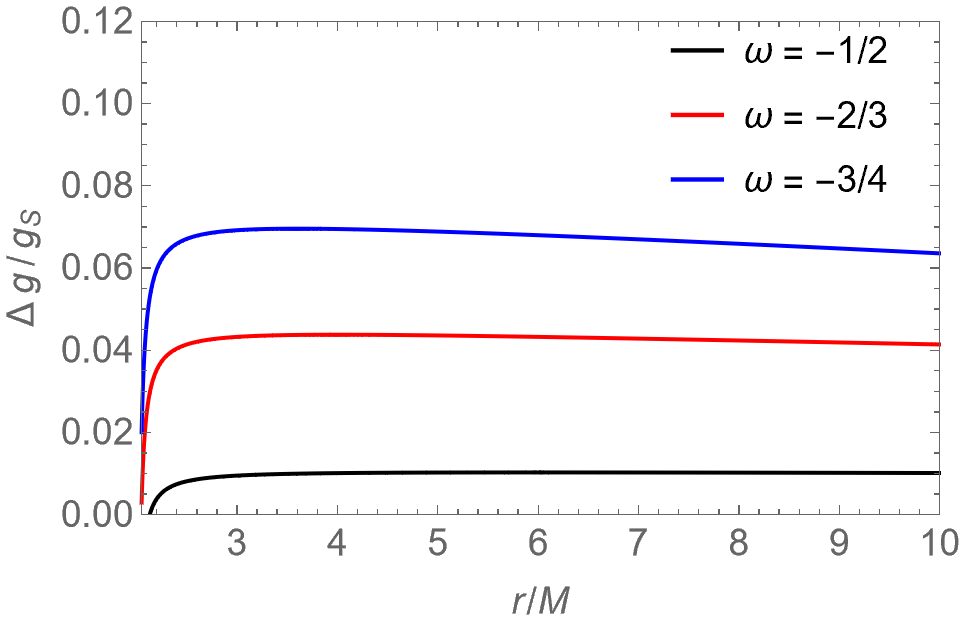}
    \caption{Difference in the redshift factor $g$ for static spherical emission, $\Delta g = g - g_{\rm S}$, between the
quintessence–corrected Schwarzschild spacetime and the standard Schwarzschild case.  The static geodesic observer is located at the radius where $f(r)$ attains its maximum value: For $\omega = -1/2$, $r_{\rm obs} \approx 252M$; for $\omega = -2/3$, $r_{\rm obs} \approx 45M$; and for $\omega = -3/4$, $r_{\rm obs} \approx 27M$; for Schwarzchild case, $r_{\rm obs} \to \infty$.  For all quintessence–corrected Schwarzschild spacetimes, the normalization constant is fixed to $c=0.001$.  }
    \label{fig:g_difference}
\end{figure}

\begin{figure}
    \centering
    \includegraphics[width=0.9\linewidth]{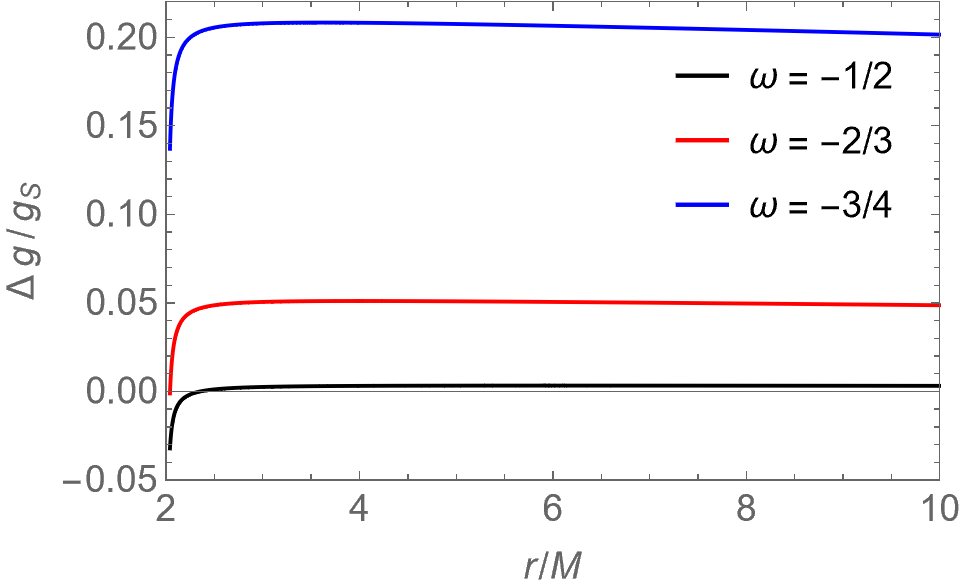}
    \caption{Same as Fig.~\ref{fig:g_difference}, but now with the location of the static observer fixed at $r_{\rm obs}=100M$ for all spacetimes, including the Schwarzschild case.}
    \label{fig:g_difference2}
\end{figure}

For static spherical emission, we have the four-velocity of the emitter
\begin{equation}
    u_\text{em}^{\alpha}=\left(\frac{1}{\sqrt{f(r)}},0,0,0\right)\,.
\end{equation}
The four-velocity of a static observer at $r_{\rm obs}$
\begin{equation}
    u_\text{obs}^{\beta}=\left(\frac{1}{\sqrt{f(r_\text{obs})}},0,0,0\right).
\end{equation}
and the redshift factor $g$ is given by Eq.~\eqref{eq:redshift}
\begin{equation}
    g=\sqrt{\frac{f(r)}{f(r_\text{obs})}}.
\end{equation}
For Schwarzschild case, it is natural to take $r_{\rm obs} \to \infty$ and we have $g_{\rm S} = \sqrt{f(r)}$.  In Fig.~\ref{fig:g_difference}, we show the difference in the redshift
factor for static spherical emission, $\Delta g = g - g_{\rm S}$, between the
quintessence–corrected Schwarzschild spacetime and the standard Schwarzschild
geometry.  For all curves in Fig.~\ref{fig:g_difference}, the normalization constant is fixed to $c=0.001$. The location of the static geodesic observer is set to $r_{\rm m}$, which depends sensitively on the equation-of-state parameter:
for $\omega = -1/2$, $r_{\rm m} \approx 252M$; for $\omega = -2/3$, $r_{\rm m} \approx 45M$; and for $\omega = -3/4$, $r_{\rm m} \approx 27M$. We observe that the deviation in the redshift factor increases as the equation-of-state parameter $\omega$ becomes more negative. Moreover, over essentially the entire radial range shown in the figure, the redshift factor in the quintessence–modified spacetime remains consistently larger than in the Schwarzschild case. This trend contrasts with the behaviour seen in Fig.~\ref{jt_difference}, where the sign of the deviation for the emission coefficient changes with radius. A similar feature is also observed in Fig.~\ref{fig:g_difference2}, where the location of the static observer is fixed at $r_{\rm obs}=100M$ for all spacetimes, including the Schwarzschild case.

The observed intensity for the static emissions is governed by
\begin{equation}
    I_\text{static}=\pm \int_\text{ray} \frac{rj_\text{t}(r)g^3}{\sqrt{f(r_\text{obs})}\sqrt{r^2-b^2f(r)}} \mathrm{d}r\label{eq:istatic}\,,
\end{equation}
where the positive sign corresponds to light rays falling into the black hole, and the negative sign corresponds to those moving away from it. For light rays not captured by the black hole, their trajectory exhibits a ``turning point'' at $r_\text{turn}$, which satisfies the equation $r_\text{turn}^2=b^2f(r_\text{turn})$.

\begin{figure*}[htbp]
    \centering
    \includegraphics[width=0.48\linewidth]{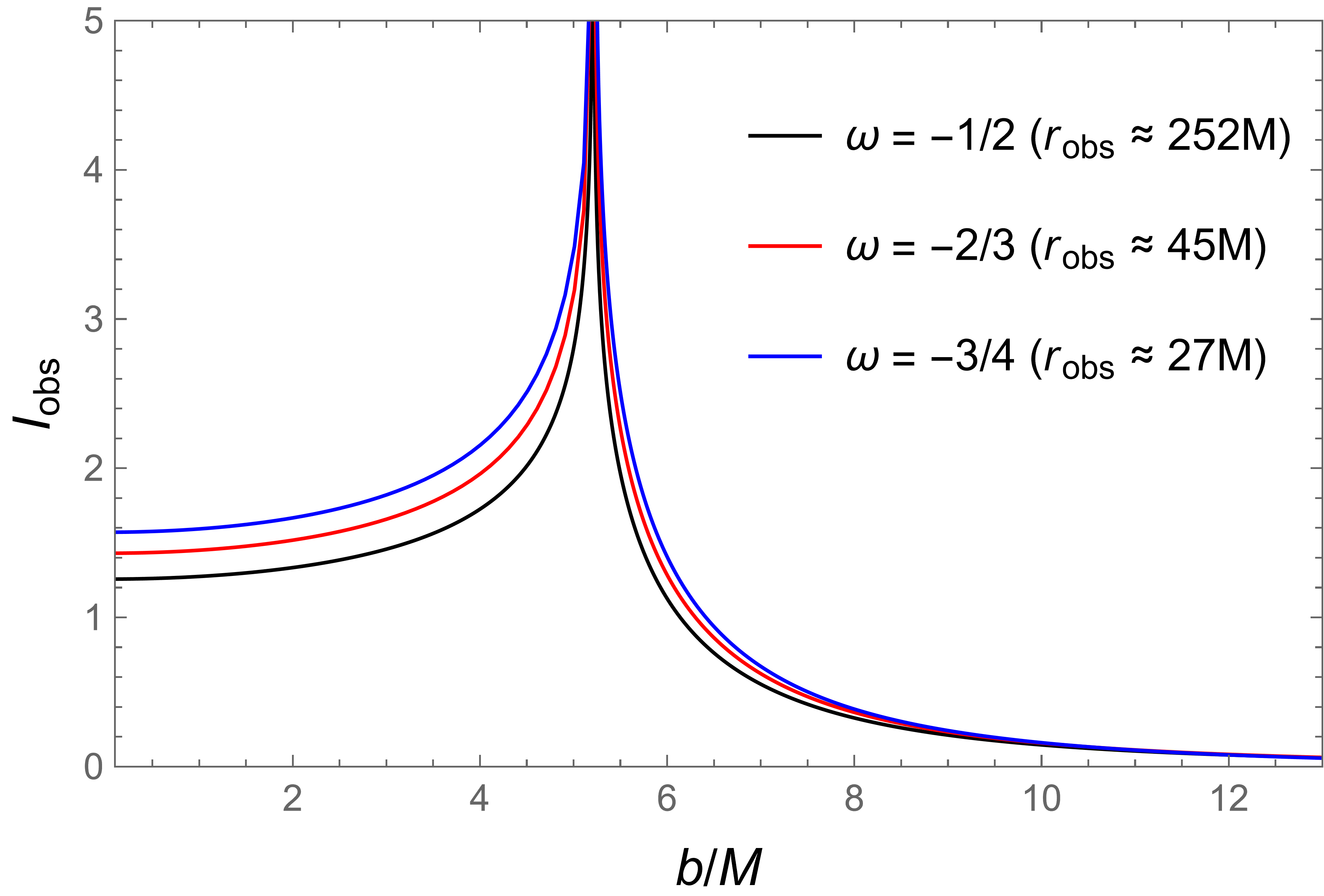}
    \hfill
    \includegraphics[width=0.48\linewidth]{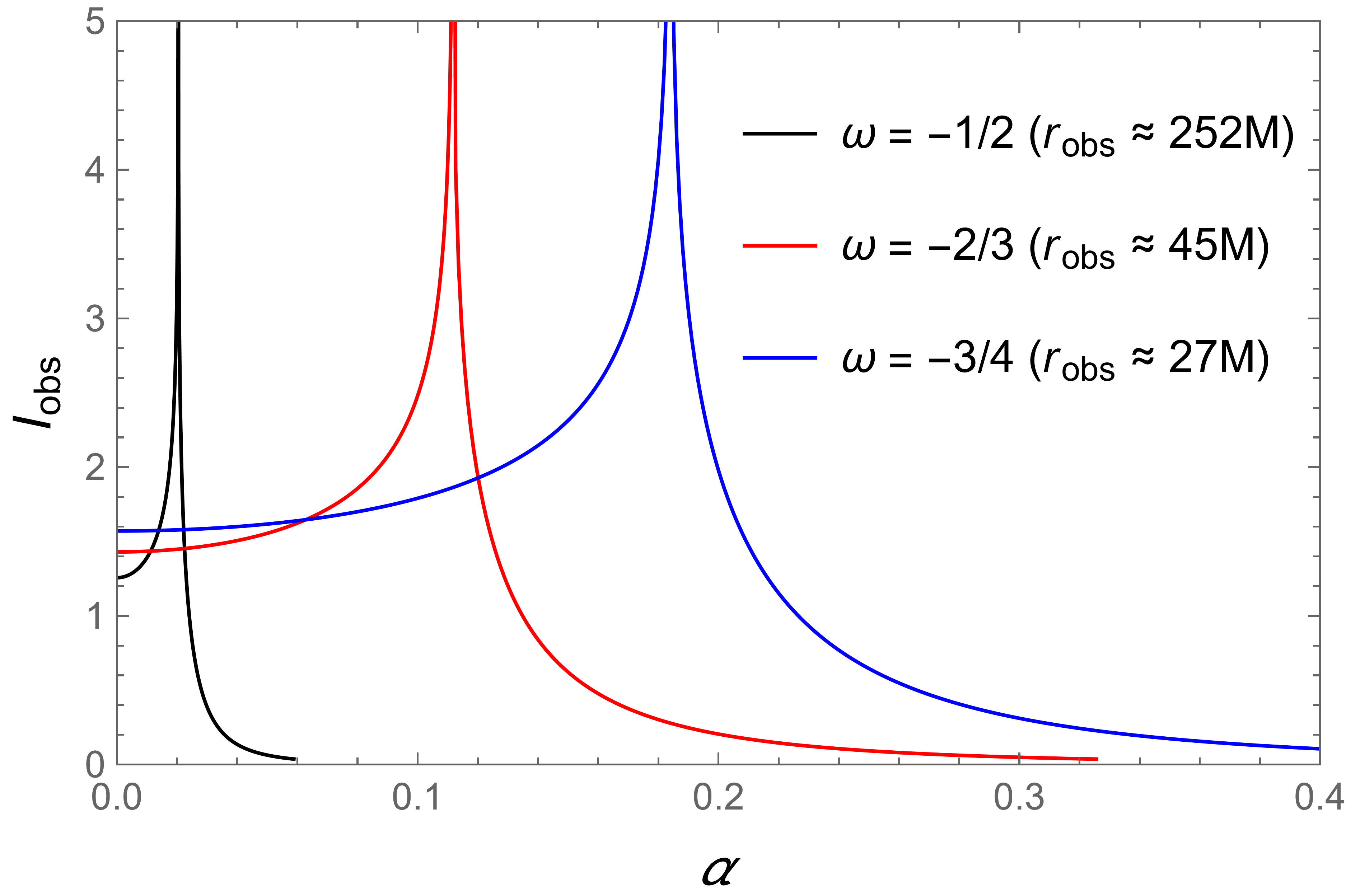}
    \caption{Observed intensity $I_{\rm obs}$ of static spherical emission measured by a static observer located at $r_{\rm obs}=r_{\rm m}$, where the four-acceleration of static observers vanishes. Note that $r_{\rm obs}$ differs for different values of $\omega$, since $r_{\rm m}$ itself depends on $\omega$. 
The left panel shows $I_{\rm obs}$ as a function of the impact parameter $b$, while the right panel displays $I_{\rm obs}$ as a function of the apparent angle $\alpha$. 
For different values of $\omega$, the peak of the observed intensity is always located near $b=b_{\rm c}\simeq 5.2M$, corresponding to the photon sphere. 
However, the corresponding apparent angular size of this peak differs for different $\omega$. For all quintessence–corrected Schwarzschild spacetimes, the normalization constant is fixed to $c=0.001$.}
    \label{fig:obsstatic}
\end{figure*}

\begin{figure*}[htbp]
    \centering
    \includegraphics[width=0.48\linewidth]{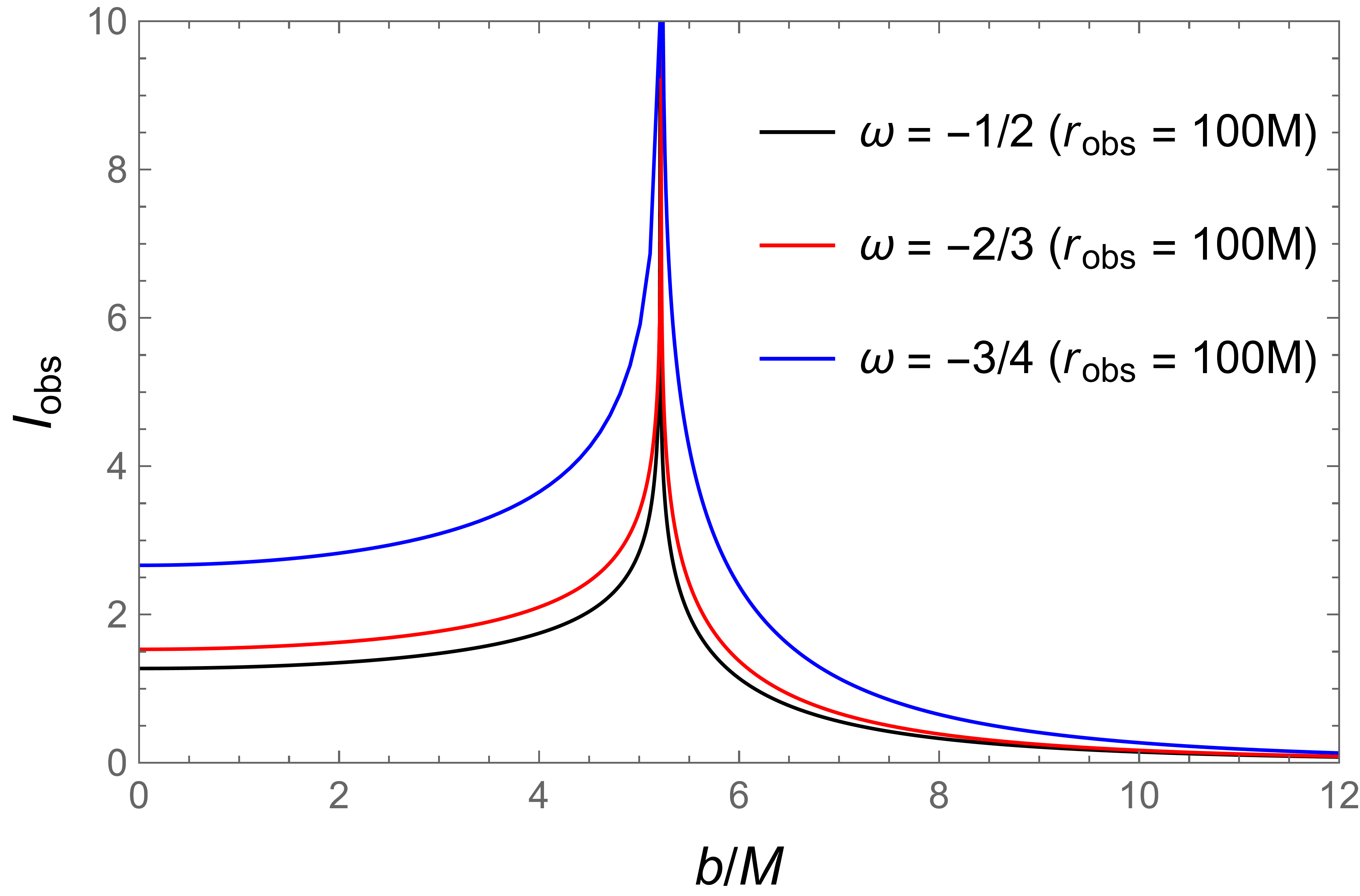}
    \hfill
    \includegraphics[width=0.48\linewidth]{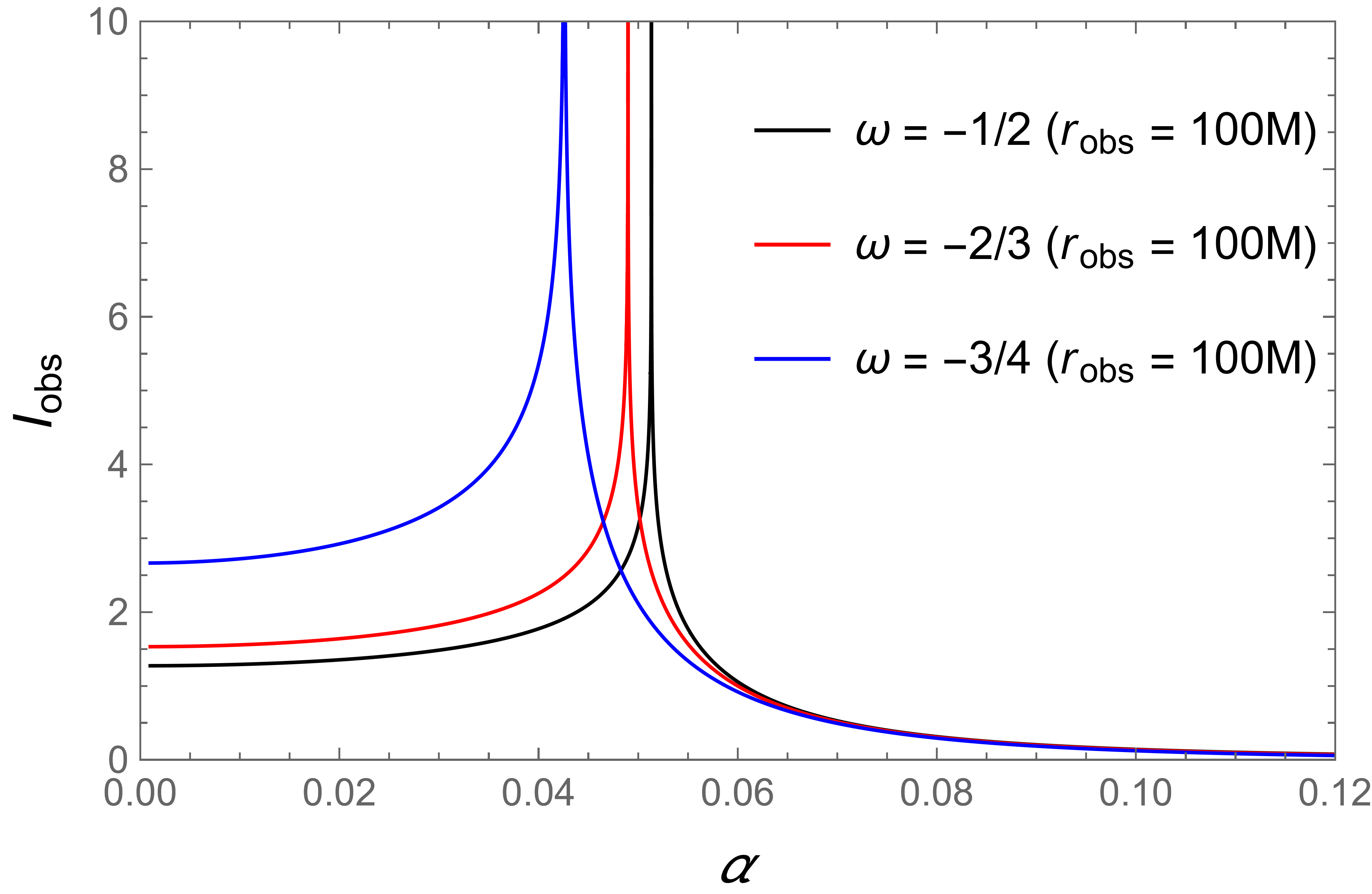}
    \caption{Same as Fig.~\ref{fig:obsstatic}, but with the static observer located at a fixed radius $r_{\rm obs}=100M$ for all spacetimes.}
    \label{fig:obsstatic100M}
\end{figure*}

To illustrate the role of the observer’s location in shaping the observed intensity profile, we present in Fig.~\ref{fig:obsstatic} and Fig.~\ref{fig:obsstatic100M} the bolometric intensity measured by static observers placed at two different radii. 
In Fig.~\ref{fig:obsstatic}, the observer is located at $r_{\rm obs}=r_{\rm m}$, where the metric function $f(r)$ attains its maximum and the four-acceleration of static observers vanishes. 
In Fig.~\ref{fig:obsstatic100M}, by contrast, the observer is placed at a fixed radius $r_{\rm obs}=100M$ for all spacetimes, representing a more conventional static observer far from the black hole.

As shown in the left panel of Fig.~\ref{fig:obsstatic}, the intensity profiles exhibit a pronounced peak near the critical impact parameter $b=b_{\rm c}$, corresponding to photons asymptotically approaching the photon sphere. 
This feature is independent of the observer’s location and is consistent with the values of the critical impact parameter $b_{\rm c}$ listed in Table~\ref{tab:comparison}, as expected for the transition between captured and escaping photon trajectories. 
Notably, while the peak appears at nearly the same value of $b$ for different values of $\omega$, the corresponding apparent angular position of the peak differs, as shown in the right panel of the figure. When the observer is placed instead at a fixed large radius $r_{\rm obs}=100M$, as shown in Fig.~\ref{fig:obsstatic100M}, the qualitative features of the intensity distribution remain similar. 
This behavior reflects the dependence of the locally measured angle $\alpha$ on both the spacetime geometry and the observer’s location, and highlights the fact that, in non-asymptotically flat spacetimes, the impact parameter alone does not fully characterize the observable angular structure.

Finally, we observe that in both Fig.~\ref{fig:obsstatic} and Fig.~\ref{fig:obsstatic100M}, the overall brightness of the accretion flow increases as the equation-of-state parameter $\omega$ becomes more negative.
This trend can be attributed mainly to the enhanced gravitational blueshift-compared to the Schwarzschild case-associated with more negative values of $\omega$, as illustrated in Figs.~\ref{fig:g_difference} and~\ref{fig:g_difference2}.

\subsection{Infalling Spherical Emission}

\begin{figure*}[htbp]
    \centering
    \includegraphics[width=0.48\linewidth]{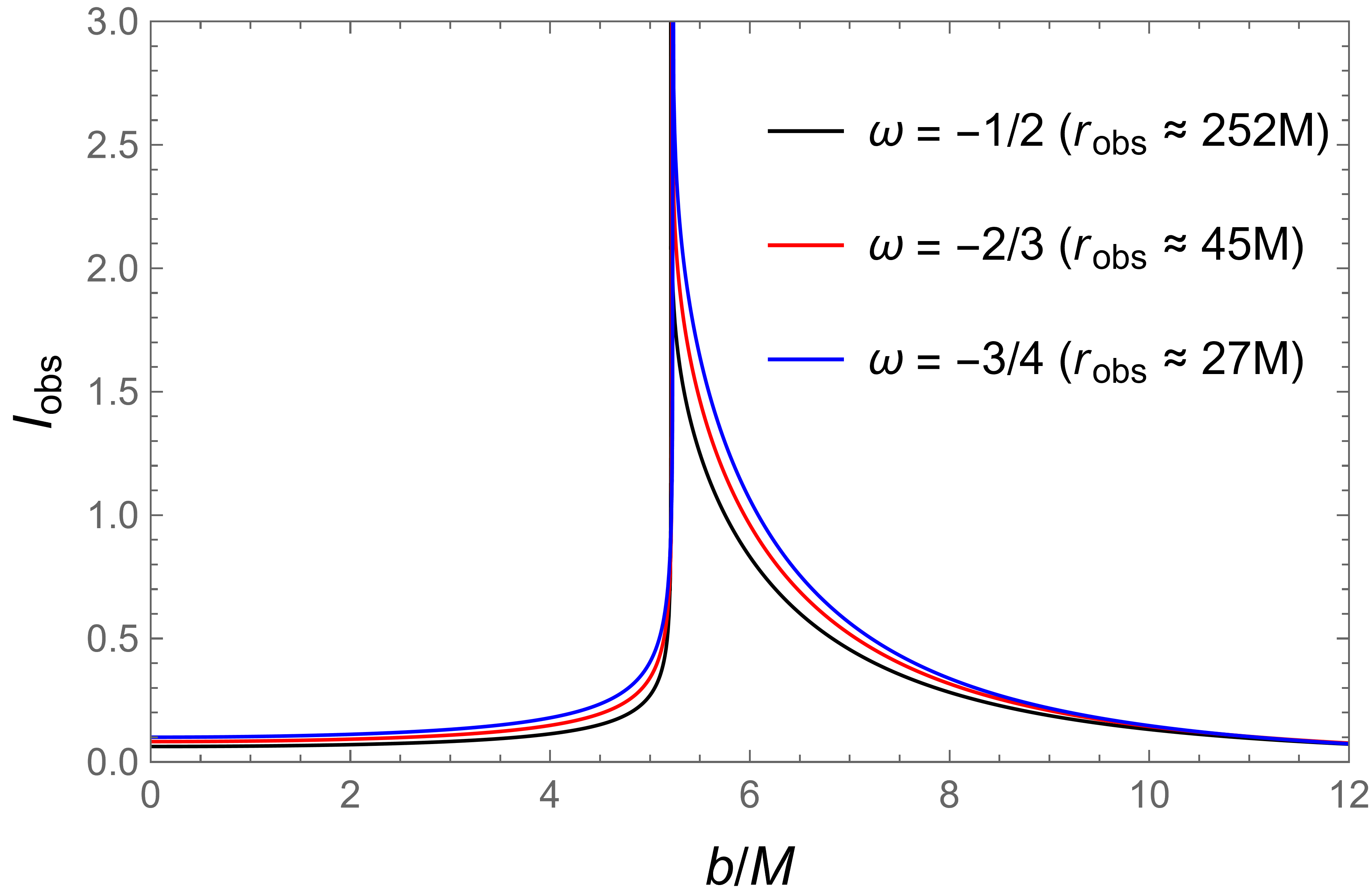}
    \hfill
    \includegraphics[width=0.48\linewidth]{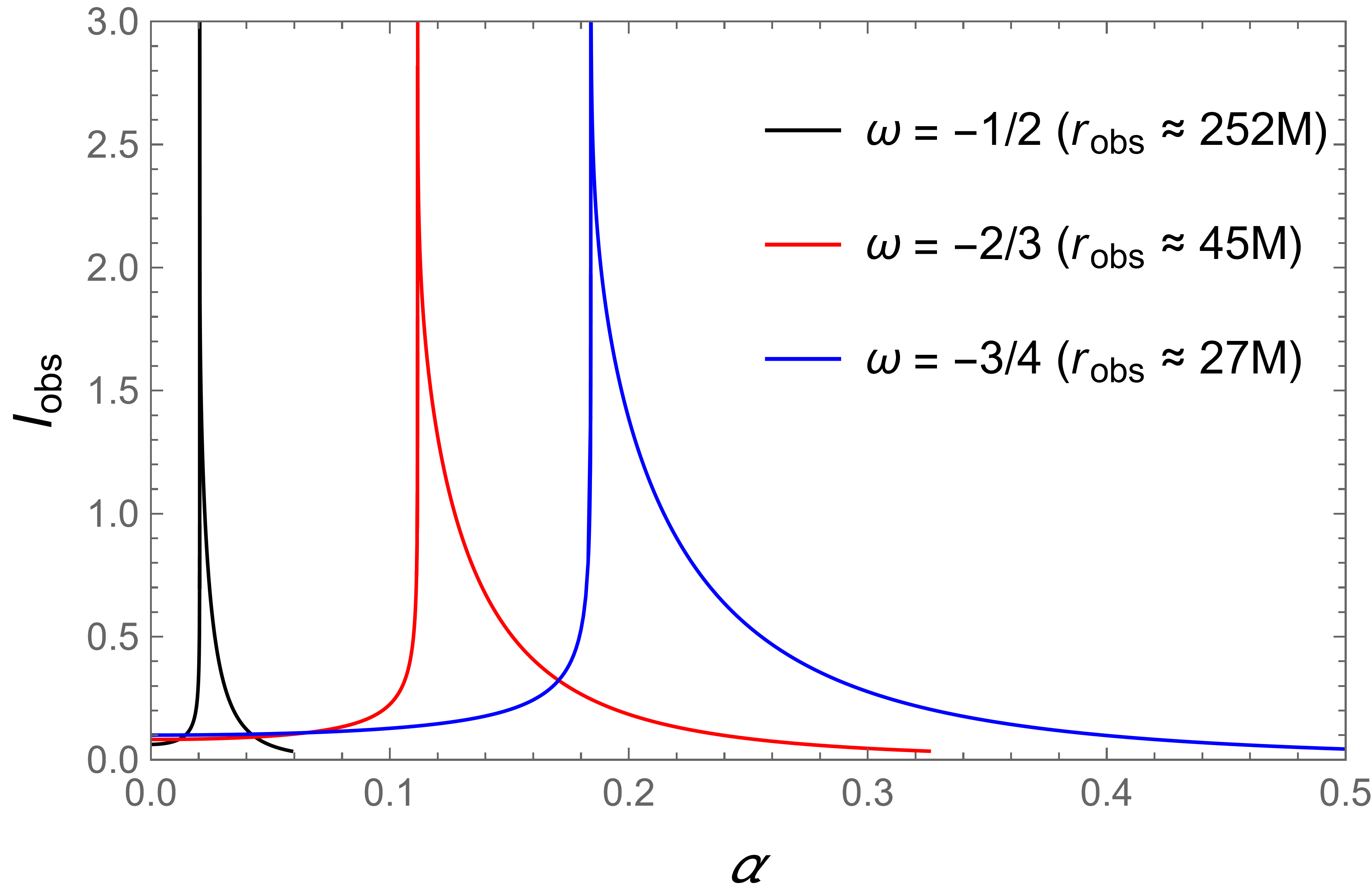}
    \caption{Observed intensity $I_{\rm obs}$ of infalling spherical emission measured by a static observer located at $r_{\rm obs}=r_{\rm m}$.
    The left panel shows $I_{\rm obs}$ as a function of the impact parameter $b$, while the right panel displays $I_{\rm obs}$ as a function of the apparent angle $\alpha$.
    For all quintessence-corrected Schwarzschild spacetimes, the normalization constant is fixed to $c=0.001$.}
    \label{fig:obsinfall}
\end{figure*}

We now turn to the case of infalling spherical emission. 
The emitting gas is assumed to be freely infalling radially from rest at $r=r_{\rm m}$, so that its four-velocity takes the form
\begin{equation}
    u_{\rm em}^{\alpha}
    =
    \left(
    \frac{\sqrt{f(r_{\rm m})}}{f(r)},
    -\sqrt{f(r_{\rm m})-f(r)},
    0,
    0
    \right).
\end{equation}
This choice corresponds to geodesic motion in the radial direction and provides a natural generalization of the static emission model considered previously.

For this infalling emitter, the redshift factor $g$, defined as the ratio between the photon frequency measured by the observer and that measured in the emitter’s rest frame, is given by
\begin{equation}
    g
    =
    \frac{
    r f(r) / \sqrt{f(r_{\rm obs})}
    }{
    r \sqrt{f(r_{\rm m})}
    \pm
    \sqrt{f(r_{\rm m})-f(r)} \,
    \sqrt{r^2-b^2 f(r)}
    } .
\end{equation}
Here the sign corresponds to outgoing or ingoing photon trajectories along the ray.

The observed intensity associated with infalling spherical emission can then be written as
\begin{equation}
    I_{\rm infalling}
    =
    \pm
    \int_{\rm ray}
    \frac{
    r \, j_{\rm t}(r) \, g^3
    }{
    \sqrt{f(r_{\rm obs})}
    \sqrt{r^2-b^2 f(r)}
    }
    \, \mathrm{d}r ,
\end{equation}
where, as before, we have neglected the absorption, corresponding to the optically thin regime.

The corresponding intensity profiles are shown in Fig.~\ref{fig:obsinfall}. 
Overall, the qualitative behavior closely resembles that of the static emission case. 
However, for $b<b_{\rm c}$, the observed intensity is systematically suppressed. 
This reduction reflects the enhanced redshift experienced by photons emitted by infalling gas, which includes both gravitational redshift and the additional Doppler shift associated with the inward motion of the emitter.

As in the static case, the peak of the observed intensity consistently appears near the critical impact parameter $b=b_{\rm c}$ for different values of $\omega$, confirming the central role of the photon sphere in shaping the observed signal. 
Nevertheless, the apparent angular position of this peak depends sensitively on $\omega$, once again emphasizing that, in non-asymptotically flat spacetimes, the impact parameter alone is insufficient to fully characterize the observable angular structure.

\section{Shadow Observed by a Geodesic (free-falling) Observer}
\label{sec:sec4}

As discussed in Sec.~\ref{sec:sec3}, in the Schwarzschild spacetime a static observer located at spatial infinity provides a natural reference frame for defining the black-hole shadow, owing to the asymptotic flatness of the geometry. 
In the Schwarzschild--quintessence spacetime, however, the presence of the quintessence field renders the spacetime non-asymptotically flat, and the metric no longer approaches the Minkowski form at large radii. 
As a consequence, the notion of a static distant observer loses its global significance and can only be defined locally. 
At large distances, a more physically motivated reference frame is therefore associated with observers following timelike geodesics of the spacetime, which represent force-free motion governed solely by the background geometry.

We therefore introduce a \emph{geodesic (free-falling) observer}, whose four-velocity is determined by the timelike geodesics of the spacetime.  Such an observer is freely falling and locally inertial, and provides a natural generalization of the asymptotic static observer in non-asymptotically flat spacetimes.

In this section, we focus on the apparent size of the photon ring, which determines the overall extent of the black-hole shadow, as already discussed in Sec.~\ref{sec:sec3}. 
To this end, we construct an orthonormal tetrad adapted to the geodesic observer, derive the photon four-momentum components measured in this local frame, and compute the apparent shape of the black-hole shadow as perceived by such an observer.

Assuming that the free-falling observer has a four-velocity $v^\mu$, 
the normalization condition $v^\mu v_\mu = -1$ implies that its components can be written as
\begin{align}\label{eq:4-v}
    v^\mu = \left( \frac{E_o}{f(r)},\ \pm\sqrt{E_o^2 - f(r)},\ 0,\ 0 \right)\,,
\end{align}
where $E_o$ denotes the conserved (specific) energy of the observer along the timelike geodesic.  The upper (lower) sign corresponds to a freely outgoing (freely infalling) observer.

In the standard Schwarzschild spacetime, it is customary to set the conserved energy per unit mass to unity, $E_o=1$, since this choice corresponds to an observer initially at rest at spatial infinity, where $\displaystyle \lim_{r\to\infty} f(r)=1$. 
For a quintessence-corrected black hole, however, the spacetime is generally non-asymptotically flat, and there is no preferred notion of ``rest at infinity'' to fix the normalization of $E_o$. 
A physically motivated alternative is to choose
\begin{align}
   E_o = \sqrt{f(r_{\rm m})}, 
\end{align}
which corresponds to an observer that is initially at rest at $r=r_{\rm m}$ (remind that $f(r)$ attains its maximum value at $r=r_{\rm m}$) and subsequently follows a freely falling (inward or outward) timelike geodesic. 

It should be emphasized that the choice of $E_o$ is not unique. 
For non-geodesic observers, $E_o$ is not
necessarily conserved and may even depend explicitly on $r$. 
As will be discussed later, a particular choice such as $E_o = 1 - 2M/r$ is relevant for a specific family of observers in the Schwarzschild--de~Sitter spacetime. 
In the present analysis, however, we restrict ourselves to geodesic observers with constant $E_o$. 
With this choice, Eq.~\eqref{eq:4-v} defines the timelike basis vector $e_{0}{}^{\mu}=v^\mu$ of the orthonormal tetrad associated with the free-falling observer.

To complete the tetrad, we construct the remaining three spacelike unit vectors that are orthogonal to $e_{0}{}^{\mu}$. 
A convenient choice is
\begin{align}
e_{1}{}^{\mu} &= \left( \pm\frac{\sqrt{E_o^2 - f(r)}}{f(r)},\ E_o,\ 0,\ 0 \right)\,, \\
e_{2}{}^{\mu} &= \left( 0,\ 0,\ \frac{1}{r},\ 0 \right)\,, \\
e_{3}{}^{\mu} &= \left( 0,\ 0,\ 0,\ \frac{1}{r\sin\theta} \right)\,,
\end{align}
where $e_{1}{}^{\mu}$ is constructed to be orthogonal to $e_{0}{}^{\mu}$ and normalized such that $g_{\mu\nu} e_{1}{}^{\mu} e_{1}{}^{\nu} = +1$.

\begin{figure*}[htbp]
    \centering
    \includegraphics[width=0.48\linewidth]{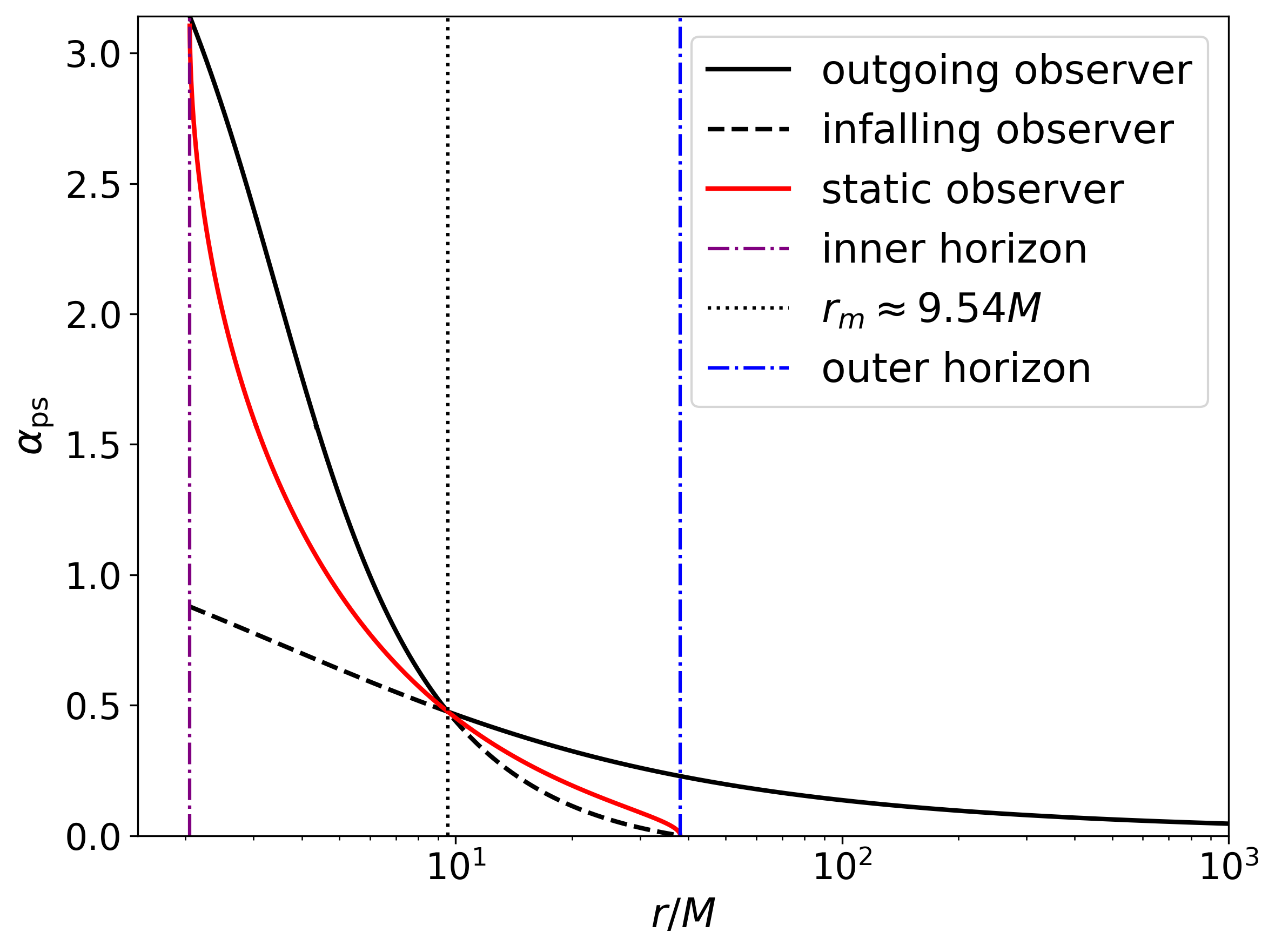}
    \label{fig:alpha_singlew}
    \hfill
    \includegraphics[width=0.48\linewidth]{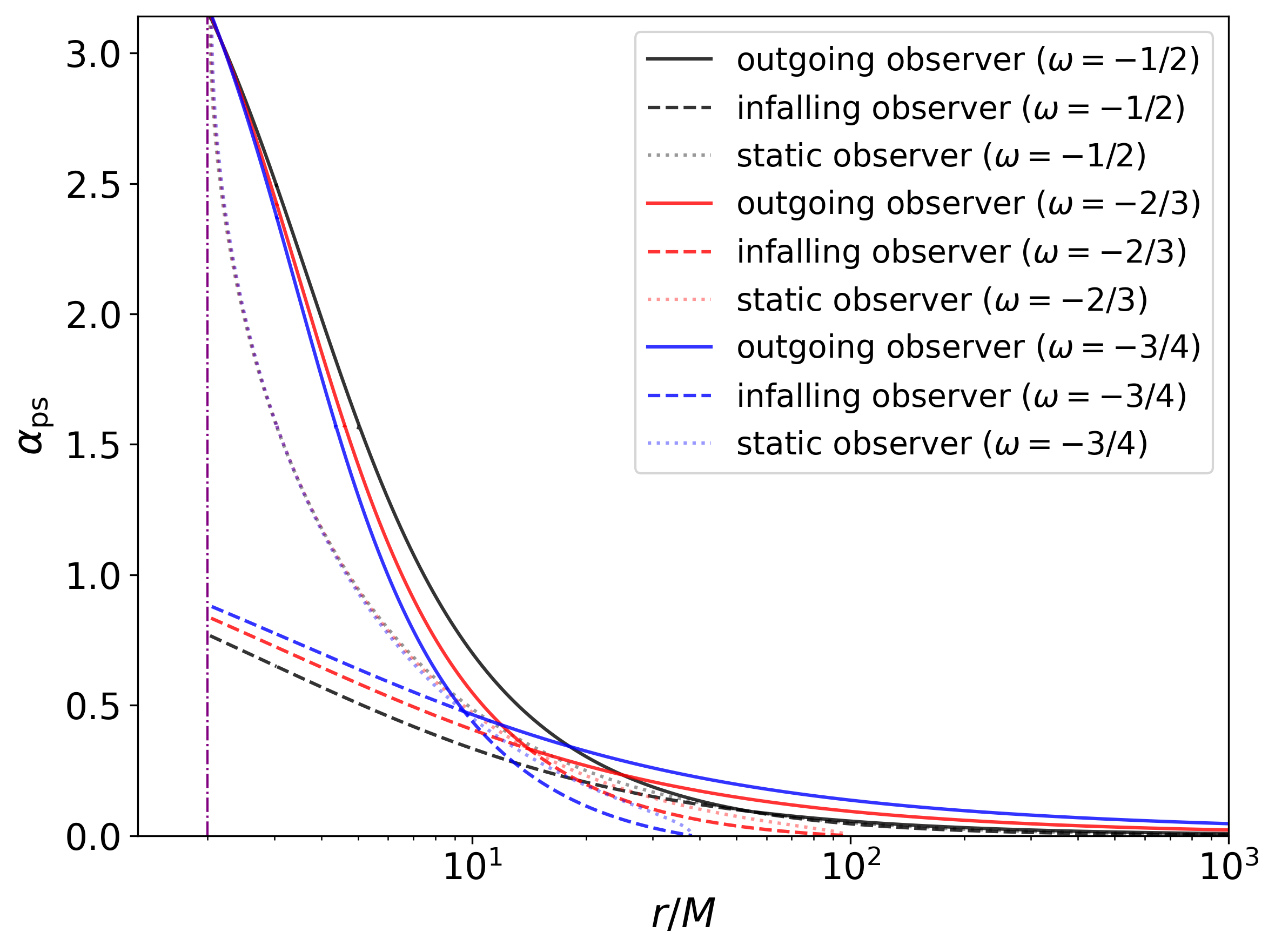}
    \label{fig:alpha_multiw}
    \caption{Angular radius of the photon sphere observed by different observers located at various radii. 
Left panel: the case with $c=0.01$ and $\omega=-3/4$. For these parameters, the spacetime contains two horizons: the inner (black-hole) horizon at $r_{\rm h}\approx 2.05M$ and the outer horizon at $r_{\rm out}\approx 38.13M$. The metric function $f(r)$ attains its maximum at $r_{\rm m}\approx 9.54M$. 
Right panel: results for different values of $\omega$, all with $c=0.01$. For these parameters, the inner horizon radii are all close to the Schwarzschild value. Note that the dotted curves for different $\omega$ nearly overlap for static observers at $r<10M$.}
    \label{fig:alpha_geodesic_observer}
\end{figure*}

Using the identity~\cite{Carroll2004}
\begin{align}
    e^{a}{}_{\mu} = \eta^{ac} g_{\mu\nu} e_{c}{}^{\nu}\,,
\end{align}
the corresponding dual tetrads are obtained as
\begin{align}
    e^{0}{}_{\mu} &= \left( E_o,\ \mp\frac{\sqrt{E_o^2 - f(r)}}{f(r)},\ 0,\ 0 \right)\,, \\
    e^{1}{}_{\mu} &= \left( \mp\sqrt{E_o^2 - f(r)},\ \frac{E_o}{f(r)},\ 0,\ 0 \right)\,, \\
    e^{2}{}_{\mu} &= \left( 0,\ 0,\ r,\ 0 \right)\,, \\
    e^{3}{}_{\mu} &= \left( 0,\ 0,\ 0,\ r\sin\theta \right)\,.
\end{align}
Therefore, the angle radius observed (i.e., $\alpha$) in this local frame satisfies
\begin{align}
    \tan \alpha &=\mp\frac{\dd y'}{\dd x'} \bigg|_{r_{\rm obs}} \notag \\
    &= \mp \frac{e^{3}{}_{3} \dd \phi}{e^{1}{}_{0} \dd t+ e^{1}{}_{1}\dd r}\bigg|_{r_{\rm obs}}\,,
\end{align}
which leads to 
\begin{align}
    \tan ^2 \alpha &= \frac{b^2 f^2}{r^2}\left(E_o\sqrt{1-\frac{b^2 f}{r^2}}\mp\sqrt{E_o^2-f}\right)^{-2}\Bigg|_{r_{\rm obs}} \,,\\ \label{eq:angle_expandO}
    \sin ^2 \alpha &=\frac{b^2 f^2}{b^2 f^2+r^2\left(E_o\sqrt{1-\frac{b^2f}{r^2}}\mp\sqrt{E_o^2-f}\right)^{2}}\Bigg|_{r_{\rm obs}}\,.
\end{align}
The up sign in Eq.~~\eqref{eq:angle_expandO} represents the free outfalling geodesic observer, while the lower sign represents the free infalling geodesic observer. Even though the result is obtained initially in the region of the domain of outer communication,  it can also be applied to the region outside the outer horizon for free outfalling geodesic observer. Fig.~\ref{fig:alpha_geodesic_observer} illustrates how the apparent angular radius of the photon sphere (taking $b=b_c$ in Eq.~~\eqref{eq:angle_expandO}) depends on the observer’s radial position for both freely outgoing and freely infalling geodesic observers. For comparison, the angular radius measured by a static observer at the same radius, given by Eq.~\eqref{eq:angle}, is also shown.   As shown in the left panel of Fig.~\ref{fig:alpha_geodesic_observer} ($c=0.01$ and $\omega=-3/4$), the two families of observers perceive markedly different angular sizes for almost all observer positions.  The only exception occurs at $r_{\rm obs}=r_{\rm m}$, where both the freely outgoing and freely infalling observers are momentarily at rest (though unstable) with respect to the static frame, and therefore measure the same angular radius.

For the freely outgoing observer, as the radial position increases from the 
black-hole horizon to the outer horizon and further outward, the apparent angular radius $\alpha$ decreases monotonically from $\pi$ to a finite nonvanishing value at the outer horizon. In contrast,  for the freely ingoing observer, $\alpha$ starts from a value smaller than $\pi$ at the black-hole horizon and decreases continuously to zero as the observer approaches the outer horizon.

Moreover, we find that the angular radius 
measured by a freely outgoing observer is always larger than that measured by the corresponding static observer at the same radius, whereas the freely infalling observer always measures a smaller angular radius compared to the static observer. These trends are fully consistent with the relativistic aberration of light: when an observer moves \emph{toward} the black hole (infalling), incoming photon directions are aberrated toward the forward direction of motion, effectively concentrating them (compared with the corresponding static observer at the same radius) and thereby making the apparent angular size of the photon sphere smaller. Conversely, for an observer moving \emph{away} from the black hole (outgoing), the aberration acts in the opposite manner, shifting photon directions toward the backward hemisphere and stretching the angular size on the sky, so that the photon sphere appears larger than for a static observer.

From the right panel of Fig.~\ref{fig:alpha_geodesic_observer}, we observe that, for different values of $\omega$, although the angular radius measured by a static observer is nearly identical at small radii, the angular radius measured by a freely falling observer is clearly distinguishable. 
As $|\omega|$ increases, the angular radius of the photon sphere becomes smaller at small radii, while it becomes larger at larger radii.

We now turn to the asymptotic behaviour of Eq.~\eqref{eq:angle_expandO} in the limit
$r\to\infty$, which corresponds to the angular radius of the photon sphere as measured
by a freely outfalling observer located at a very large distance.  
For $-1<\omega<-1/3$, the exponent $3\omega+1$ is negative, such that 
$r^{-(3\omega+1)}$ is a positive power of $r$.  
Consequently,
\begin{equation}
    f^{2}(r) \;\longrightarrow\; c^{2}\, r^{-(6\omega+2)} ,
    \qquad (r\to\infty),
\end{equation}
and Eq.~\eqref{eq:angle_expandO} asymptotically yields
\begin{equation}\label{eq:sinalpha_limit}
    \sin^{2}\alpha_{\rm ps}
    \;\longrightarrow\;
    \frac{c\,b_{c}^{\,2}}
         {c\,b_{c}^{\,2}+r^{\,3+3\omega}},
    \qquad (r\to\infty).
\end{equation}

Since $3+3\omega>0$ for any $\omega>-1$, the denominator in
Eq.~\eqref{eq:sinalpha_limit} diverges as $r\to\infty$, implying that
the angular radius of the photon sphere measured by a freely outfalling observer
asymptotically vanishes:
\begin{equation}
    \alpha_{\rm ps}\;\longrightarrow\;0,
    \qquad (r\to\infty,\; -1<\omega<-1/3).
\end{equation}

The only exceptional case is the limit $\omega=-1$, namely the
Schwarzschild--de~Sitter spacetime.  
In this case $3+3\omega=0$ and Eq.~\eqref{eq:sinalpha_limit} reduces to a finite, nonzero value,
\begin{equation}
    \alpha_{\rm ps}
    \;\longrightarrow\;
    \frac{\sqrt{c}\, b_{c}}{M},
\end{equation}
where we have reinstated the mass scale $M$ so that $c$ becomes the usual
dimensionless cosmological parameter ($c\ll 1$).  
Recalling from Eq.~\eqref{eq:ps} that for $\omega=-1$ the photon-sphere radius is
$r_{\rm ps}=3M$, we obtain
\begin{equation}
    b_{c} \simeq 3\sqrt{3}\, M ,
    \qquad (c\ll 1),
\end{equation}
and therefore
\begin{equation}
    \alpha_{\rm ps}
    \;\longrightarrow\;
    3\sqrt{3c},
    \qquad (r\to\infty,\; \omega=-1).
\end{equation}

It is noteworthy that this finite asymptotic angle agrees with the result obtained
in Ref.~\cite{Perlick:2018iye}, where a comoving observer approaching infinity in Schwarzschild--de~Sitter spacetime measures the same nonvanishing angular radius. This agreement suggests that an outgoing geodesic observer at infinity is effectively equivalent to a comoving observer at infinity in Schwarzschild--de~Sitter spacetime.
Indeed, Eq.~\eqref{eq:4-v} reproduces the four-velocity of the comoving observer when the energy parameter is chosen as $E_o = 1 - 2M/r$. With this choice, one can explicitly verify that our general expression Eq.~\eqref{eq:angle_expandO} reduces to Eq.~(36) of Ref.~\cite{Perlick:2018iye}. 

\begin{figure}
    \centering
    \includegraphics[width=1\linewidth]{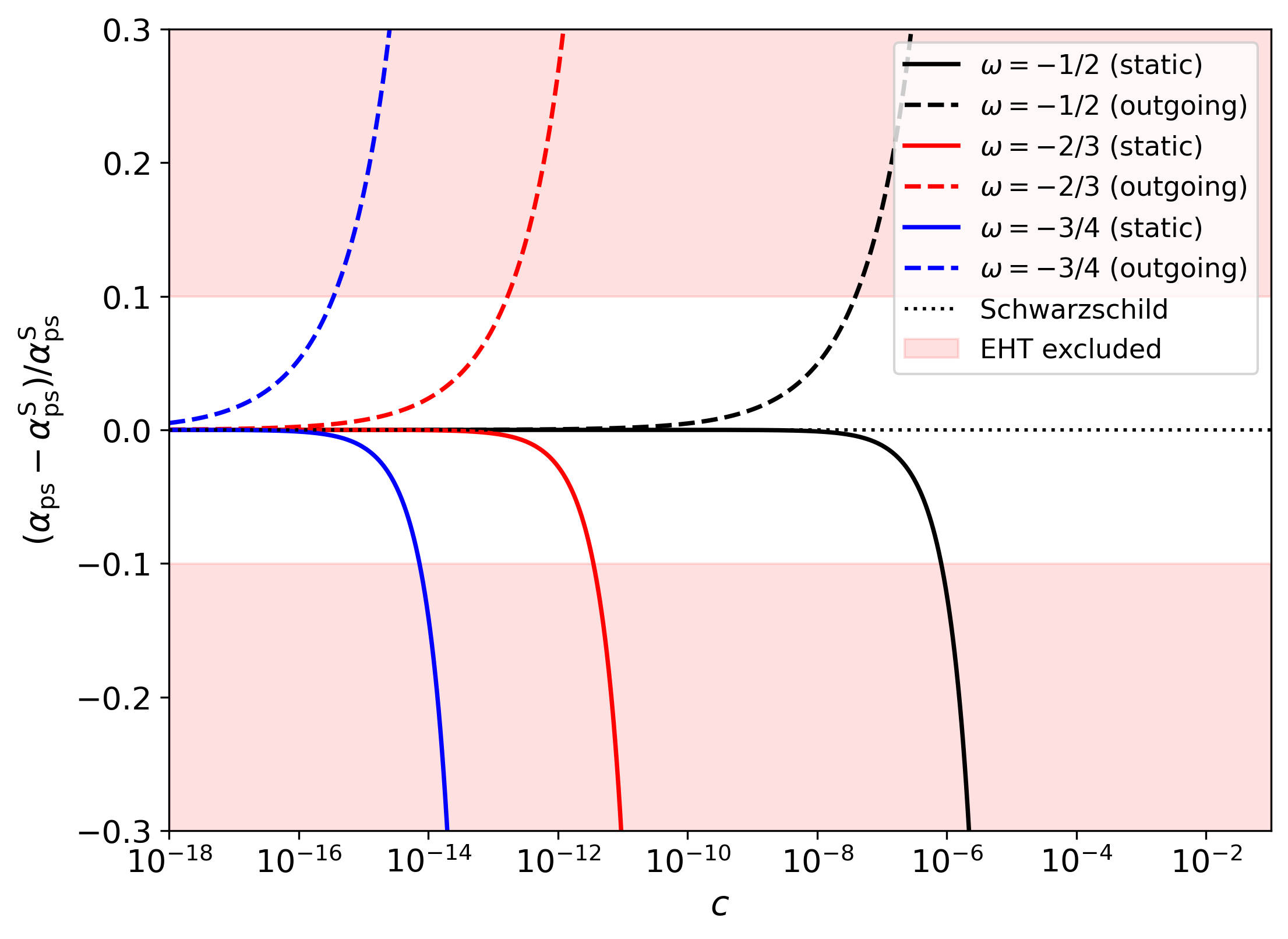}
\caption{Relative deviation of the angular radius of the photon sphere from the Schwarzschild value,
$(\alpha-\alpha_{\rm S})/\alpha_{\rm S}$, as a function of the dimensionless quintessence parameter $c$.
The curves are obtained assuming that M87$^\ast$ is described by a quintessence black hole
with mass $M = 6.5\times10^{9} M_\odot$ and observer distance $D = 16.8\,\mathrm{Mpc}$.
Solid curves correspond to a static observer, while dashed curves correspond to an outgoing geodesic observer
with energy normalization $E_o=\sqrt{f(r_{\rm m})}$.
Different colors indicate different values of the equation-of-state parameter $\omega$.
The shaded red region is excluded by the EHT observations at the $\pm10\%$ level.
}
\label{fig:EHT_static_vs_deodesic}
\end{figure}

Fig.~\ref{fig:EHT_static_vs_deodesic} shows the relative deviation of the photon-sphere angular radius
from the Schwarzschild prediction as a function of the quintessence parameter $c$,
assuming that M87$^\ast$ is modeled as a quintessence black hole with
$M = 6.5\times10^{9} M_\odot$ and $D = 16.8\,\mathrm{Mpc}$.
Solid curves correspond to a static observer, while dashed curves represent an outgoing geodesic observer
with the energy normalization fixed by $E_o=\sqrt{f(r_{\rm m})}$.
For sufficiently small $c$, the two observer prescriptions give consistent results.
As $c$ increases, their predictions gradually deviate, especially for larger $|\omega|$.
Requiring agreement with the EHT measurement of the M87$^\ast$~\cite{EventHorizonTelescope:2019dse,EventHorizonTelescope:2019ggy} shadow at the $\pm10\%$ level
excludes the shaded region and imposes the corresponding constraints on the quintessence parameter $c$. For our choice of $E_o$, Fig.~\ref{fig:EHT_static_vs_deodesic} shows explicitly that, for more negative values of the equation-of-state parameter $\omega$, the constraints imposed by EHT observations become increasingly stringent, regardless of whether the observer is static or follows an outgoing geodesic. 
This behavior arises because, at large distances, the contribution of the quintessence term to the spacetime geometry becomes more significant as $\omega$ decreases.

\section{Summary and conclusions }\label{Conclusion}

In this work, we have investigated the influence of a quintessence-like field on the
optical appearance and shadow properties of a Schwarzschild black hole.
Motivated by the fact that the presence of quintessence renders the spacetime
non-asymptotically flat, we have systematically examined how the black-hole shadow
depends not only on the spacetime geometry, but also on the motion and location of the observer.

We first analyzed the null and timelike geodesic structure of the
Schwarzschild--quintessence spacetime.
Using a perturbative approach, we derived approximate analytical expressions for
the event horizon, the photon-sphere radius, the innermost stable circular orbit,
and the critical impact parameter.
These results provide a transparent understanding of how the quintessence parameters
modify the spacetime geometry near the black hole.

We should emphasize that the present analysis can, in principle, be extended to a superposition of multiple quintessence components with different equations of state.
Such a generalization, however, would mainly introduce additional parameters without qualitatively altering the underlying physical mechanisms discussed here, and is therefore left for future work.

We then studied the optical appearance of the black hole under spherical accretion, considering both static and infalling emission models. Special emphasis was placed on the distinction between the impact parameter and the locally measured angular radius, which becomes essential in non-asymptotically flat spacetimes. We showed that, while the photon ring consistently appears near the critical impact parameter, its apparent angular size depends sensitively on the observer’s location and state of motion.

To account for the lack of a preferred static observer at large distances, we introduced geodesic (freely falling) observers as physically motivated reference frames. We demonstrated that relativistic aberration leads to systematic differences between static and geodesic observers: infalling observers measure a smaller angular radius, whereas outgoing observers perceive an enlarged photon sphere. In the asymptotic region, the apparent angular radius vanishes for $-1<\omega<-1/3$, while it approaches a finite value in the Schwarzschild--de~Sitter limit, in agreement with previous results.

Finally, we confronted our theoretical predictions with the EHT observation of the M87$^\ast$ black-hole shadow. Requiring consistency at the $\pm10\%$ level, we derived constraints on the quintessence parameter $c$. We found that more negative values of the equation-of-state parameter $\omega$ lead to increasingly stringent constraints, and that the difference between static and geodesic observers becomes significant when the quintessence contribution is no longer negligible.

Our results underscore the necessity of carefully addressing observer dependence in the interpretation of black-hole shadow observations within non-asymptotically flat spacetimes. Moreover, unlike the Schwarzschild case, the critical impact parameter alone is insufficient to fully characterize the observed angular structure in such spacetimes. The framework developed here can be straightforwardly extended to rotating black holes or more general dark-energy-inspired modifications of gravity~\cite{Toshmatov:2015npp,Abdujabbarov:2015pqp,Ghosh:2015ovj}, providing a useful tool for future high-resolution black-hole imaging.

\acknowledgments
 This work was supported by the National Natural Science Foundation of China (Grant No. 12405063).
	
\vspace{10mm}

\bibliographystyle{jhep.bst}
\bibliography{references.bib}

\end{document}